\documentclass[a4paper]{article}
\usepackage[english]{babel}
\usepackage[utf8x]{inputenc}
\usepackage{float}
\usepackage[titletoc]{appendix}
\usepackage[a4paper,top=3cm,bottom=2cm,left=3cm,right=3cm,marginparwidth=1.75cm]{geometry}
\usepackage{amsmath,authblk}
\usepackage{graphicx,blindtext}
\usepackage[colorinlistoftodos]{todonotes}
\usepackage{array}
\usepackage[colorlinks=false]{hyperref}
\begin{document}
	\title{ \textbf{A note on electromagnetic and gravitational perturbations of the Bardeen de Sitter black hole: quasinormal modes and greybody factors}}\vspace{3cm}
	
	\author[1]{Sahel Dey\thanks{Email: saheldey917@gmail.com}}
	\author[2]{Sayan Chakrabarti\thanks{Email: sayan.chakrabarti@iitg.ac.in}}
	\affil[1]{Joint Astronomy Programme and Department of Physics, Indian Institute of Science, Bangalore-560012, India}
	\affil[2] {Department of Physics, Indian Institute of Technology Guwahati, Guwahati-781039, India}
	\date{}
	
	\maketitle
	
	
	\begin{abstract}
	 Bardeen de-Sitter (BdS) black hole is a spherically symmetric solution of Einstein's equation which is coupled to nonlinear electromagnetic field in a way that one gets a regular solution, devoid of any singularity at the origin. We compute the quasinormal (QN) frequencies for BdS black hole due to electromagnetic and gravitational perturbations. We analyse the behaviour of both real and imaginary parts of BdS QN frequencies by varying the black hole parameters and compare frequencies with Reissner-Nordstr\"{o}m de-Sitter (RN-dS) black hole. Interestingly, we find that the response of BdS and RN-dS black holes under electromagnetic and gravitational perturbations  are different when the charge parameter is varied, which can be used to understand nonlinear and linear electromagnetic fields in curved spacetime separately. A study on the dynamics of perturbation as well as the scattering from the BdS black holes using WKB approach is performed. Greybody factors and their variations with black hole parameters are investigated.
	\end{abstract}\vspace{1cm}
	
	\section{Introduction}

	It is very well known that general relativity is a theory which is plagued with the appearance of singularities. The invariant scalar curvature which necessarily tells about the gravitational field strength diverges at those spacetime singularities. Gravitational singularities appear in general relativity in the context of black holes. Black holes are objects which have singularities at the origin hidden by the event horizons. However, appearance of singularities in a theory means that the theory breaks down at the point where the singularity is present. Hence, the task of avoiding the singularities in general theory of relativity is one of the most fundamental ones and a set of solutions known as ``regular black holes'' play an important role in this context. As the name suggests, when the black hole does not have a  spacetime singularity at the origin, it is termed as a ``regular black hole''. Bardeen \cite{bardeen} obtained the first solution of regular black holes with non-singular geometry satisfying the weak energy condition. The solutions is known as the Bardeen black hole in the literature. The solution Bardeen obtained was not a vacuum solution rather gravity was modified by introducing some form of matter. Therefore an energy momentum tensor was introduced in the Einstein's equation in order to achieve that goal. The introduction of the energy momentum tensor was done in an {\it ad hoc} manner and hence the Bardeen solution lacked physical motivation. After a long time, Ay\'{o}n-Beato and Garc\'{i}a \cite{abg1} showed that the energy momentum tensor necessary to obtain regular black hole solution is essentially the gravitational field of some magnetic monopole arising out of a specific form of non-linear electrodynamics. Many other solutions \cite{abg2}-\cite{tosh1}, motivating  the avoidance of singularity was proposed thereafter. Stability properties \cite{ms1, tosh2, tosh3} and quasinormal modes \cite{nino, fern1}, thermodynamics \cite{man} and geodesic structure \cite{zcw} of such regular black holes were studied in detail. On another front, Fernando \cite{fern2} has recently found out a de Sitter branch for the regular Bardeen black hole and corresponding grey body factors for such a black hole were calculated. The stability analysis and quasinormal modes due to scalar and Fermionic perturbations were also studied \cite{wadbor} for this background.  The motivations for studying regular black holes in de Sitter space comes from the fact that our universe looks like asymptotically de Sitter at very early and late times. Observational data also indicates that our universe is going through a phase of accelerated expansion \cite{perl, riess, tegmark_sdss}, which, along with many other explanations also indicates the existence of a positive cosmological constant. Hence, the study of black holes and its various features in de Sitter space is by itself an increasingly demanding area of research. In continuation of our earlier work \cite{wadbor}, we will study the gravitational and electromagnetic perturbations of the regular Bardeen dS black hole in this paper. 
	
	The stability of a black hole spacetime is one of the most intriguing questions that one can ask in general relativity: the answer to the question of black hole
	stability under certain perturbation can answer many questions related to the black hole itself.  The study of black hole perturbations is an active area of research and has immense effect on various important properties of black holes \cite{kk, nol, v1, kon1}. Generally one studies the evolution of a field (scalar, Fermionic, electromagnetic or gravitational) in a black hole background or in a black hole-black hole collision process in order to understand the stability of that particular black hole spacetime under the specific field perturbation. It is well know that the dynamical evolution of perturbations of a black hole background can be classified into three distinct stages, the first stage consists of an initial outburst of wave which depends completely on the initial perturbing field, the second one consists of damped oscillations, known in the literature as the quasinormal modes (QNM) whose frequencies are complex numbers. The real part of these frequencies represent the real oscillation frequency of the black hole under the perturbation and the imaginary part represents damping. The final stage is a power law tail behaviour at very late times. QN frequencies not only provide us with the information about the stability of the black hole spacetime, they are used to determine the black hole parameters (mass, charge and angular momentum) too. Numerical simulations  depicting formation of a black holes in a gravitational collapses as well as that of collision of two black holes exclusively show that irrespective of the nature of the perturbations, the black hole's response will be dominated by the QNMs \cite{anninos}. One important aspect of studying black hole stability is the fact that equations governing the black hole perturbations in most of the cases can be cast into a Schr\"{o}dinger like equation. The QNMs are solutions to that Schr\"{o}dinger like wave equation with complex frequencies for boundary conditions which are completely ingoing at the horizon and purely outgoing at asymptotic infinity (for the asymptotically flat or de Sitter black holes).  It is to be noted that apart from the fact that the QN frequencies contain important information about the black hole parameters, they were also of importance from the point of view of AdS/CFT correspondence. It has been found \cite{danny,  danny1} that QNMs in AdS space time appear naturally in the description of the dual conformal field theory on the boundary. This observation has motivated the study of QNMs towards asymptotically AdS black holes \cite{horo, card1} too. On another front, despite their classical in origin, QNMs have  been shown to provide glimpses to quantum nature of black holes \cite{hod, motl, maggiore}.  
	
	A lot of work \cite{kono3}-\cite{sm1} has been done on QNMs of scalar, electromagnetic, gravitational, 
	Dirac perturbations, decay of charged fields, asymptotic QNMs and signature of quantum gravity etc in de Sitter space. However, the regular black holes in de Sitter space is comparatively a less studied regime.  In this paper, we will try to fill up the gap in the literature by discussing the QNMs of the Bardeen de Sitter (henceforth BdS) black hole due to electromagnetic and gravitational perturbations. The plan of the paper is as follows: in the next section we give a brief discussion on the BdS black hole. In section 3 we present a discussion of WKB method for calculating the QNMs along with a
	study of the Electromagnetic QNMs of the BdS black holes. Section 4 deals with the Gravitational quasinormal modes of the BdS black hole. In section 5 we give a comparative discussion about the dynamics of the perturbations. Section 6 contains a discussion about the greybody factor and its variation with the black hole parameters. Finally, in section 7 we conclude the paper with a brief discussion on future directions. 
	
		\section{A brief discussion on BdS black hole}
	This section deals with a very brief introduction to the Bardeen de Sitter (BdS) black hole following the works in \cite{fern2}. The authors of \cite{fern2} has modified the works of \cite{abg1} to incorporate a positive cosmological constant in the action. The action therefore looks like:
	\begin{equation} \label{action}
	S=\int d^4x\sqrt{-g}\left(\frac{R-2\Lambda}{16\pi}-\frac{1}{4\pi}\mathcal{L}(F)\right)
	\end{equation}
	In the above, $R$ is the Ricci Scalar and $\mathcal{L}(F)=\frac{3}{2\alpha q^2}\left(\frac{\sqrt{2q^2F}}{1+\sqrt{2q^2F}}\right)^{5/2}$ is a function of the field strength $F$ of the non-linear electrodynamics. Here field strength($F$) is defined as $F = \frac{1}{4}F^{\mu \nu}F_{\mu \nu} $ where $F_{\mu\nu}=2(\nabla_{\mu} A_{\nu}-\nabla_{\nu}A_{\mu})$. 
	The parameter $\alpha$ in $\mathcal{L}(F)$ is related to the magnetic charge ($q$) and the mass ($M$) of the space time as follows: $\alpha=\frac{q}{2M}$. The equations of motion from the above action comes out to be \cite{fern2}:
	\begin{align}
	G_{\mu\nu}+\Lambda g_{\mu\nu}&=2\left(\frac{\partial \mathcal{L}(F)}{\partial F}F_{\mu\lambda}F^{\lambda}_{\nu}-g_{\mu\nu}\mathcal{L}(F)\right)& \\
	\nabla_{\mu}\left(\frac{\partial \mathcal{L}(F)}{\partial F}F^{\nu\mu}\right)&=0\\
	\nabla_{\mu}(* F^{\nu\mu})&=0
	\end{align}
	A static spherically symmetric solution for the above set of equations exist \cite{fern2}:
	\begin{equation}
	ds^2=-\left(1-\frac{2Mr^2}{(r^2+q^2)^{3/2}}-\frac{\Lambda r^2}{3}\right)dt^2+\left(1-\frac{2Mr^2}{(r^2+q^2)^{3/2}}-\frac{\Lambda r^2}{3}\right)^{-1}dr^2+r^2(d\theta^2+\sin^2\theta d\phi^2)\label{metric}
	\end{equation}
	The zeros of the function $f(r)=1-\frac{2Mr^2}{(r^2+q^2)^{3/2}}-\frac{\Lambda r^2}{3}$ gives the horizon. The BdS black hole there can have at most three horizons corresponding to three real roots of the function $f(r)$: the black hole inner$(r_i)$ and outer horizons$(r_h)$ along with the cosmological horizon$(r_c)$.  It is to be noted that the BdS black hole is structurally similar to the Reissner-Nordstr\"{o}m-de Sitter (RNdS) or Born-Infeld de Sitter (BIdS) black holes which also admits a possibility of three distinct horizons as well as a single or degenerate horizons too (corresponding to extremal case). However, the event horizon is much larger for RNdS black hole as compared to a BdS one \cite{fern2}. The non-singular structure of the BdS geometry can be checked by direct calculation of the scalar curvatures $R$, $R_{\mu\nu}R^{\mu\nu}$, $R_{\mu\nu\lambda\sigma}R^{\mu\nu\lambda\sigma}$, which are finite everywhere as compared to divergences at $r=0$ in case of Einstein black holes except the electromagnetic field invariant $F$ which is singular at $r=0$ \cite{fern2}.  

    \section{Electromagnetic perturbations and QNMs of the BdS black hole}
	In this paper we focus our attention on the behaviour of the dynamical response of the spherically symmetric regular black hole in de Sitter space under electromagnetic and gravitational perturbations. In this section we will be discussing the electromagnetic field perturbations of the BdS black hole in order to study the behaviour of the QNMs in this background by varying a set of black hole parameters. Since our system is an open one, the black hole, after a small perturbation, relaxes to its equilibrium state by losing energy by emitting electromagnetic or gravitational radiation, depending on the nature of the underlying perturbations.  
	
	As discussed in Section 2,  BdS background metric is given by  Eqn.(\ref{metric}). Now we decompose 4-vector potential of the electromagnetic field in two parts. One is unperturbed background potential ($\bar {A_{\mu}}$) and another is perturbed part ($\delta A_{\mu}$).\\
	\begin{equation}\label{ATotal}
	A_{\mu}= \bar{A}_{\mu} + \delta A_{\mu}
	\end{equation}
	 In static and spherically symmetric background, ansatz for unperturbed 4-vector potential of magnetically charged black hole is given by 
	\begin{equation} 
	\bar{A}_{\mu}= -q {\cos\theta \delta ^{\phi}_{\mu}} \label{backg}
	\end{equation}
	 Considering spherically symmetric BdS background, perturbation in vector potential can be written as a superposition of vector spherical harmonics, where $Y_{\ell m} (\theta, \phi)$ are standard scalar spherical harmonics:
\begin{eqnarray}\nonumber
	{\delta A}_{\mu_{\rm{axial}}}=\displaystyle\sum_{\ell=0}^{\infty} \displaystyle\sum_{m=-\ell}^{\ell}
	\begin{bmatrix}
	0       \\
	\\
	0       \\
	\\
	a_0(t,r)\frac{1}{\sin\theta}\frac{\partial Y_{\ell m}}{\partial\phi}\\ 
	\\ 
	-a_0(t,r)\sin\theta \frac{\partial Y_{\ell m}}{\partial\theta}       
	\end{bmatrix},~ 
	{\delta A}_{\mu_{\rm{polar}}}=	\displaystyle\sum_{\ell=0}^{\infty} \displaystyle\sum_{m=-\ell}^{\ell} \begin{bmatrix}
	    a_1(t,r)Y_{\ell m}       \\
		\\
		a_2(t,r)Y_{\ell m}       \\
		\\
		a_3(t,r)\frac{\partial Y_{\ell m}}{\partial\theta}\\
		\\  
		a_3(t,r) \frac{\partial Y_{\ell m}}{\partial\phi}
	\end{bmatrix}
	\end{eqnarray}
	 It is well known that, under the angular space inversion transformation $(\theta,\phi) \rightarrow  (\pi-\theta,\pi+\phi) $, first part of the transformation changes sign as $(-1)^{(\ell +1)}$ termed as axial or odd part and second part changes sign as $(-1)^{(\ell)}$ termed as polar or even part. As $\delta A_{\mu}$ is decoupled under parity transformation, we have only focused on the axial modes which is the first part of perturbed potential. 
	The Electromagnetic (EM) field tensor is defined in terms of the general 4-vector potential as follows :
	\begin{equation}
	F_{\mu\nu}= 2\left(\nabla_{\mu} A_{\nu}-\nabla_{\nu} A_{\mu}\right)
	\end{equation}
	and generalized Maxwell's equations for nonlinear electrodynamics are represented by
	\begin{equation}
	\nabla_{\nu} \left(\mathcal{L}_{F} F^{\mu\nu} \right) =0 \label{feq}
	\end{equation}
	Where $ \mathcal{L}_{F} = \frac{\partial \mathcal{L}(F) }{\partial F} .$ Considering only background field($\bar{A}_{\mu}$), non vanishing terms of EM field tensor are $F_{\theta\phi} = -F_{\phi\theta} =q \sin\theta $
	 and strength(F) of the EM field becomes $ 2q^{2}/r^{4}. $\\ Taking into account the perturbation part ($\delta A_{\mu}$) along with background 4-potential, non zero field tensor components are as follows :
	\begin{eqnarray}
	\begin{split} 
    F_{t\theta} & = \frac{1}{\sin\theta} \frac{\partial a_{0} }{\partial t} 
     \frac{\partial Y_{\ell m} }{\partial \phi}\\ 
     F_{t\phi} & =  -\sin\theta \frac{\partial a_{0} }{\partial t}
     \frac{\partial Y_{\ell m} }{\partial \theta} \\
      F_{r\theta} & =  \frac{1}{\sin\theta} \frac{\partial a_{0} }{\partial r}
      \frac{\partial Y_{\ell m} }{\partial \phi}\\
     F_{r\phi} & =  -{\sin\theta} \frac{\partial a_{0} }{\partial r}
     \frac{\partial Y_{\ell m} }{\partial \theta}\\
     F_{\theta\phi} & =  {\sin\theta}\left( q+\ell(\ell+1) a_{0}Y_{\ell m}\right)
     \end{split}   
	\end{eqnarray}
	All non zero contravariant components of EM field tensor are following:
		\begin{eqnarray}
		\begin{split}
		F^{t\theta} & = -\frac{1}{fr^{2}\sin\theta} \frac{\partial a_{0} }{\partial t}
		\frac{\partial Y_{\ell m} }{\partial \phi}\\ 
		F^{t\phi} & =  \frac{1}{fr^{2}\sin\theta} \frac{\partial a_{0} }{\partial t}
		\frac{\partial Y_{\ell m} }{\partial \theta} \\
		F^{r\theta} & =  \frac{f}{r^{2}\sin\theta} \frac{\partial a_{0} }{\partial r}
		\frac{\partial Y_{\ell m} }{\partial \phi}\\
		F^{r\phi} & =  -\frac{f}{r^{2}\sin\theta} \frac{\partial a_{0} }{\partial r}
		\frac{\partial Y_{\ell m} }{\partial \theta}\\
		F^{\theta\phi} & =  \frac{1}{r^{4}\sin\theta}\left( q+\ell(\ell+1) a_{0}Y_{\ell m}\right)
		\end{split}   
		\end{eqnarray}
For the total 4-vector potential $A_{\mu}$, field strength ($F$) remains same at zeroth-order but has components in first order, which depends on all coordinates ($t$, $r$, $\theta$ and $\phi$). At each step of our analysis, we have only considered 1st order terms in perturbation to be in linear regime. 
	   \begin{equation}	
		  F\left(\bar{A}_{\mu} + \delta A_{\mu}\right) \approx \frac{2q^{2}}{r^{4}}+\frac{4ql(l+1)a_{0}(t,r)Y_{lm}}{r^{4}}
		\end{equation}
This is the crucial point to note in eletromagnetic perturbation for electrically and magnetically charged black holes in nonlinear electrodynamics, where perturbation can not alter field strength at first order approximation. We write total field strength as $F=\bar{F}+\delta F$ where $\bar F\left(r\right)=2q^{2}/r^{4}$ and $\delta F(t,r,\theta,\phi)=\frac{4ql(l+1)a_{0}(t,r)Y_{lm}}{r^{4}}$. We expand $\mathcal{L}_{F}$ in the vicinity of $\bar{F}$ using Taylor series upto first order term. $\mathcal{L}_{F} \approx \bar{\mathcal{L}}_{\bar{F}}\left(\bar{F}\right)+
\bar{\mathcal{L}}_{\bar{F} \bar{F}} \delta F$.
 Here $\bar{\mathcal{L}}_{\bar{F}}=\frac{d\bar{\mathcal{L}}}{d\bar{F}},
\bar{\mathcal{L}}_{\bar{F} \bar{F}}=\frac{d\bar{\mathcal{L}}_{\bar{F}}}{d \bar{F}}.$ We also define $\bar{\mathcal{L}}^{'}_{\bar{F}}=\frac{d\bar{L}_{\bar{F}}}{dr}.$\\

For any free index $\mu$, Eqn.\ref{feq} becomes
\begin{equation}
\frac{\partial \left(\mathcal{L}_{F} F^{\mu t} \right) }{\partial{t}}
+\frac{1}{r^2}\frac{\partial \left(r^{2}\mathcal{L}_{F} F^{\mu r} \right) }{\partial{r}}  \label{expand_feq}
+\frac{1}{\sin\theta}\frac{\partial \left(\sin\theta\mathcal{L}_{F} F^{\mu \theta} \right) }{\partial{\theta}}
+\frac{\partial \left(\mathcal{L}_{F} F^{\mu \phi} \right) }{\partial{\phi}} = 0
\end{equation}

 For $\mu=$ $\theta$ and $\phi$, Eqn.\ref{expand_feq} simplifies to 
 
 \begin{equation}
 -\frac{\partial^{2}{a_0}}{\partial t^{2}}
 +\frac{f}{\bar{\mathcal{L}}_{\bar{F}}}\frac{\partial \left(f \bar{\mathcal{L}}_{\bar{F}}\frac{\partial a_{0}}{\partial r} \right)}{\partial r}
 +\frac{f \ell \left( \ell+1 \right)}{r^2}\left(1-\frac{4q^{2}\bar{\mathcal{L}}_{\bar{F} \bar{F}}}{r^4 {\bar{\mathcal{L}}_{\bar{F}}}}\right)a_0=0 \label{wave-no-transform}
 \end{equation} 
 
\begin{figure}[h]
		\centering
		\includegraphics[height= 6.0 cm, width=7.5 cm]{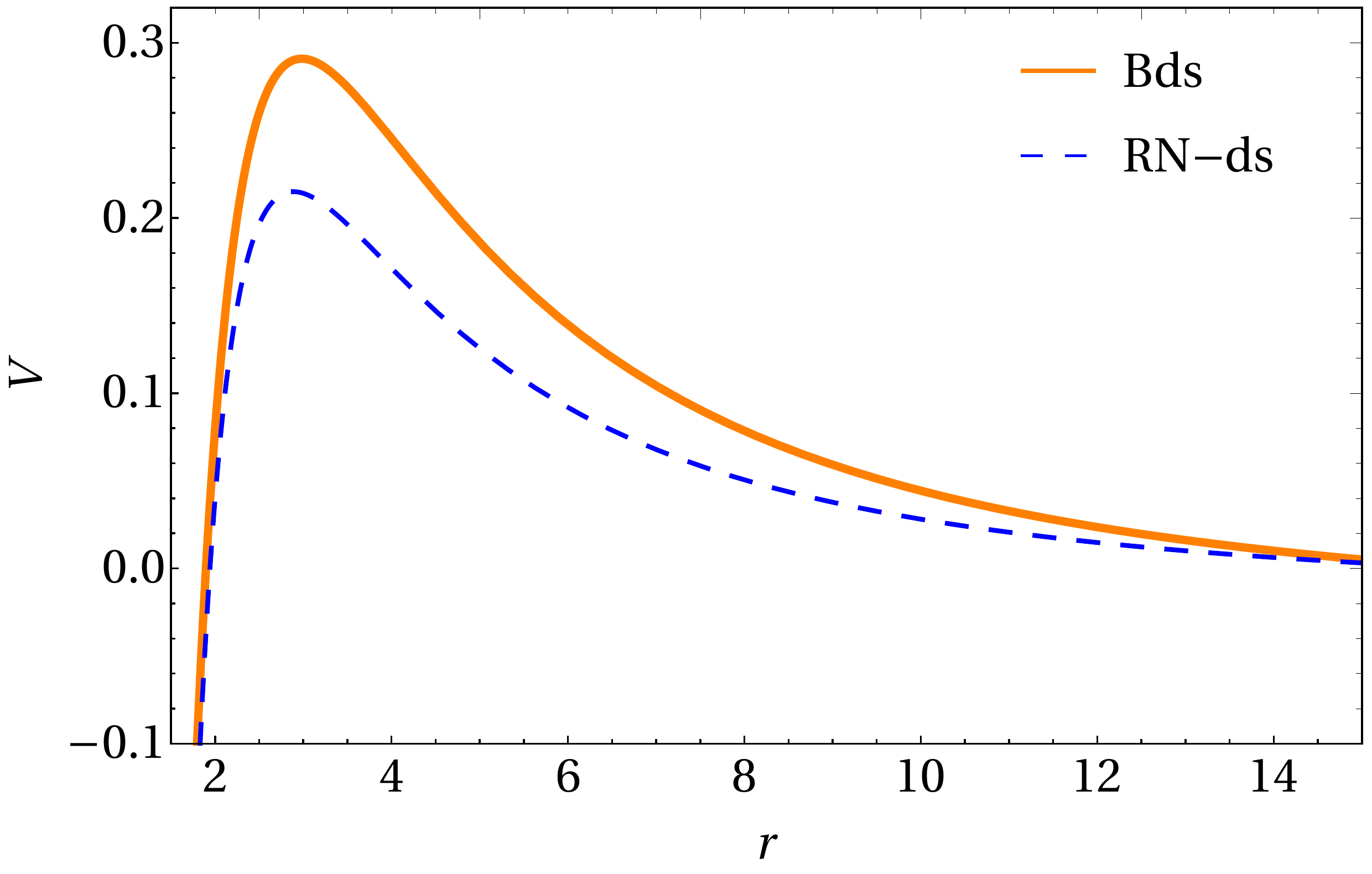}
		\caption{Effective potential $V$ for BdS and RN-dS black holes for $q=0.40$, $\ell=2$ and $\Lambda=0.01$.}\label{plotep}
	\end{figure}

 To remove first order derivative term of $a_{0}\left(t,r\right)$ from Eqn.\ref{wave-no-transform} , we use standard tortoise coordinate$(r_*)$ transformation  $dr_*=\frac{dr}{f(r)}$ and scale the variable $a_0(t,r)$ to $\Psi(t,r)=\frac{a_0(t,r)}{\sqrt{{\bar{\mathcal{L}}_{\bar{F}}}}}$. Now $\Psi$ represents solution of wave Eqn.\ref{wave-equ} with an effective potential profile $V(r)$
		\begin{equation}
	 \frac{\partial ^2 \Psi(t,r_*)}{\partial t^2}-\frac{\partial ^2 \Psi(t,r_*)}
	 {\partial r^2 _*}+\Psi(t,r_*)V(r)=0 \label{wave-equ}
	 \end{equation} 

	\begin{equation}
	V(r)=f\left[\frac{\ell(\ell+1)}{r^2}\left(1+\frac{4q^{2}\bar{\mathcal{L}}_{\bar{F} \bar{F}}}{r^4 {\bar{\mathcal{L}}_{\bar{F}}}}\right)
	-\left(\frac{f {\bar{\mathcal{L}}^{'2}_{\bar{F}}}
		-2{\bar{\mathcal{L}}_{\bar{F}}}
		\frac{ \partial \left(f \bar{\mathcal{L}}^{'}_{\bar{F}} \right)}{\partial r}
		}{4{\bar{\mathcal{L}}_{\bar{F}}}^2}\right)
	\right]\label{elpot}
	\end{equation}
	 The advantage of using the tortoise coordinate lies in the fact that the range of the coordinate now extends between $-\infty$ to $\infty$, whereas in the old radial coordinate $r$, the physically accessible region lies only between the black hole's outer horizon$(r_h)$ and the cosmological horizon$(r_c)$.
	Note also that the potential $V(r)\to 0$ as $r_{*}\to \pm\infty$.  
	 In \cite{bobir}, the authors have also computed all field components and eventually calculated the potential for electromagnetic perturbation in nonlinear electrodynamics. Apart from a few typographical errors in some of the equations (for example eqns. (36)-(38)) in that paper, the final form of the potential matches with ours in the flat space limit. In Fig.\ref{plotep}, we have examined the nature of the axial potential $V(r)$ with radial coordinate $r$ and compared BdS potental with Reissner-Nordstr\"{o}m (RN-dS) potential. For (RN-dS) black hole, $\mathcal L(F)$ linearly depends on field strength $F$ which means $\bar{\mathcal{L}}^{'}_{\bar{F}}=\bar{\mathcal{L}}_{\bar{F}\bar{F}}=0$. Importantly, it is to be noted that the overall nature of both potentials are the same, viz. (a) $V(r)$ is positive definite between the event and cosmological horizons, (b) $V(r)$ has a single maxima, which increases its height with increasing $\ell$. But for a fixed set of parameters, height of the BdS potential is larger than the RN-dS one which indicates BdS black hole has smaller absorption coefficient than RN-dS black hole. 
As already mentioned, our target in this work is to solve the wave equation with proper boundary conditions for complex QN frequencies using the sixth order WKB method developed in \cite{kon5}. It is already established in the literature that sixth order WKB method is more accurate than the third order one and the former in fact gives results coinciding with those obtained from full numerical integration of the wave equation \cite{kon5} for low overtones, i.e. for modes with small imaginary parts, and for all multipole numbers $\ell\ge 1$. The sixth order formula for a general black hole potential $V(r)$ is given by
	\begin{equation}
	\frac{i(\omega^2-V(r_0))}{\sqrt{-2V^{''}(r_0)}}-\Lambda_2-\Lambda_3-\Lambda_4-\Lambda_5-\Lambda_6 = n+\frac{1}{2}\label{qnmeqn}
	\end{equation}
	where $V(r_0)$ is peak value of $V(r)$ , $V^{''}(r_0) = \frac{d^2V}{dr_*^2}|_{r=r_0}$ , $r_0$ is the value 
	of the radial coordinate corresponding to the maximum of the potential $V(r)$ and $n$ is the overtone number. In general QN frequencies $\omega$ take the form $\omega=\omega_R- i\omega_I$, where, as mentioned earlier, the real part of $\omega$ represents actual field oscillation and imaginary part corresponds to damping of the perturbation.      
	In eqn.(\ref{qnmeqn}), $\Lambda_2$ and $\Lambda_3$ are given by \cite{will2}
    \begin{align}
    \Lambda_2&=\frac{1}{\sqrt{2V^{''}(r_0)}}\left[\frac{1}{8}\left(\frac{V_0^{(4)}}{V^{''}(r_0)}\right)(b^2+\frac{1}{4})-\frac{1}{288}\left(\frac{V_0^{(3)}}{V^{''}(r_0)}\right)^2(7+60b^2)\right]\label{part1}\\
    \Lambda_3&=\frac{(n+\frac{1}{2})}{2V^{''}(r_0)}\left[\frac{5}{6912} \left( \frac{V_0^{(3)}}{V^{''}(r_0)} \right)^4(77+188b^2) -\frac{1}{384} \left( \frac{(V_0^{(3)})^2 V_0^{(4)}}{(V^{''}(r_0))^3}\right) (51+100b^2)\right]\nonumber\\
    &+ \frac{(n+\frac{1}{2})}{2V^{''}(r_0)}\Big[\frac{1}{2304}\left( \frac{V_0^{(4)}}{V^{''}(r_0)}  \right)^2(67+68b^2)
    + \frac{1}{288}\left(\frac{V_0^{(3)}V_0^{(5)}}{(V^{''}(r_0))^2}\right)(19+28b^2)\nonumber\\
    &-\frac{1}{288} \left(\frac{V_0^{(6)}}{V^{''}(r_0)}\right)(5+4b^2)\Big]\label{part2}.
    \end{align}
	In the above expression $b=n+\frac{1}{2}$ , $V_0^{(n)}=d^nV/dr_*^n$ at $r=r_0$ and
	$\Lambda_4$, $\Lambda_5$ and $\Lambda_6$ can be found in the Appendix of \cite{kon5}. 
	The above method also works extremely well in the eikonal limit of large $\ell$ corresponding to large quality
	factors.

\begin{figure}[h]
\centering
\begin{minipage}{.5\textwidth}
 \centering
  \includegraphics[height=5 cm,width= 7 cm]{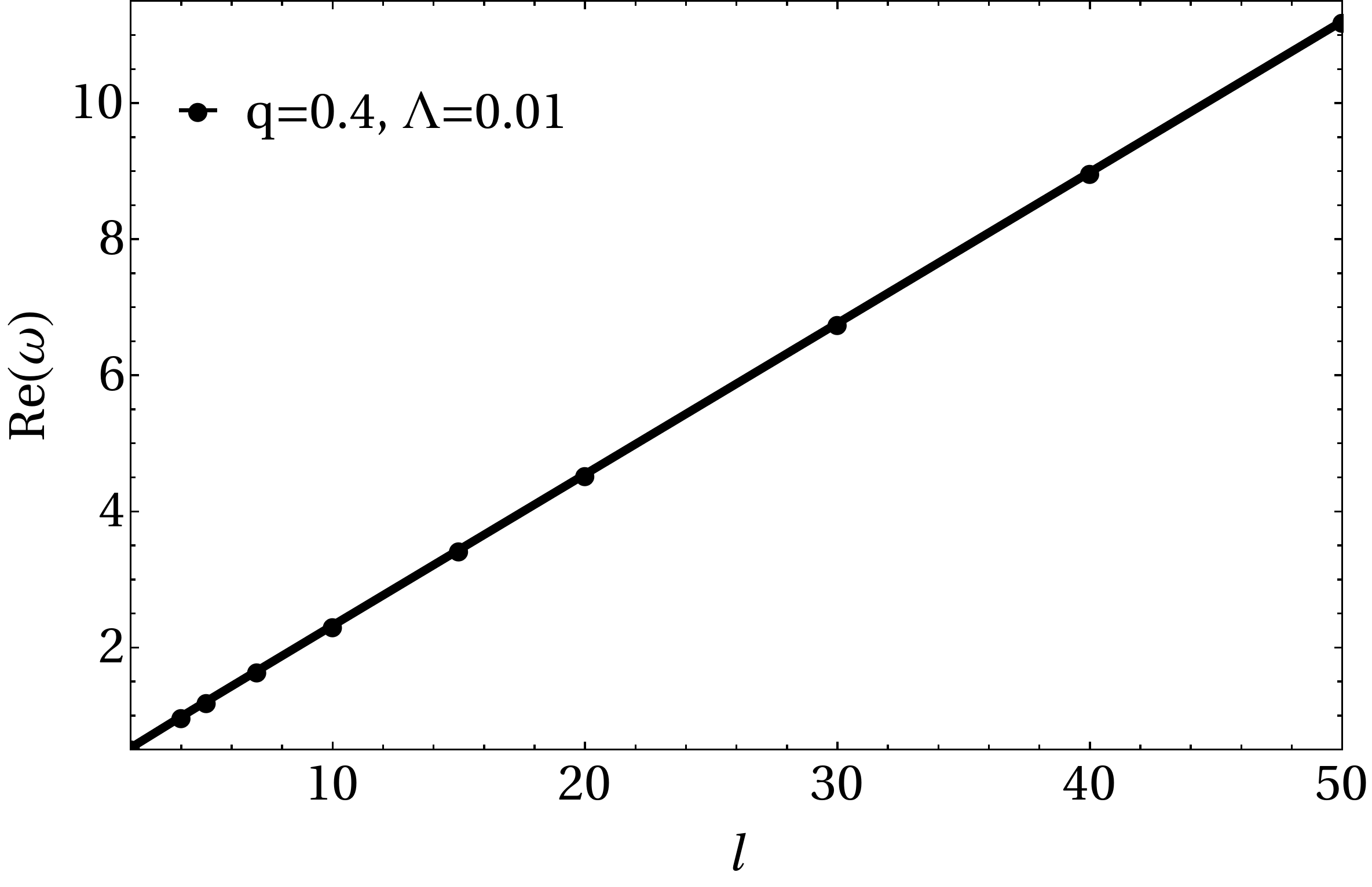}
\end{minipage}%
\begin{minipage}{.5\textwidth}
  \centering
   \includegraphics[height=5 cm,width= 7 cm]{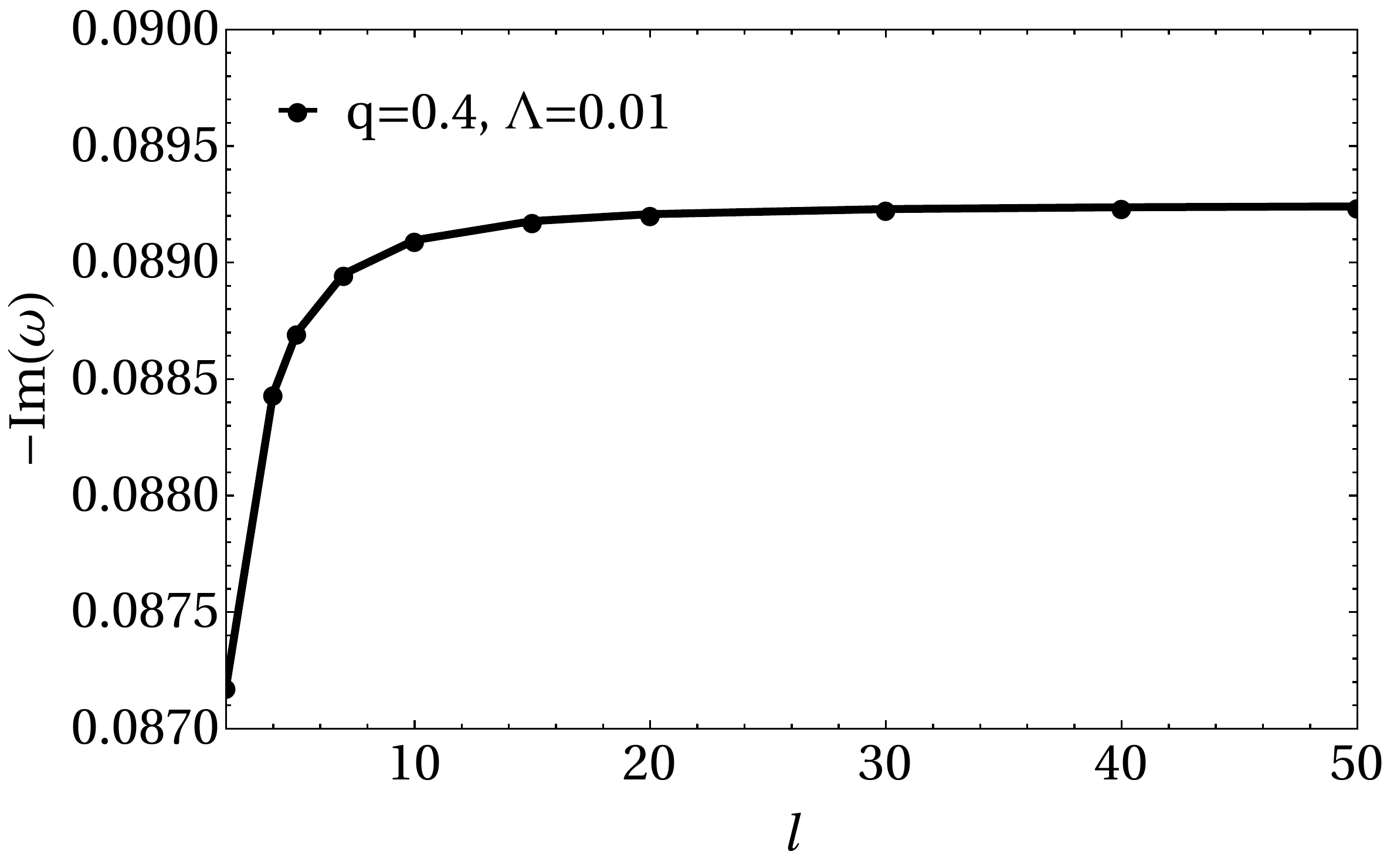}
\end{minipage}
       \caption{Variation of Re $\omega$ and -Im $\omega$  with multipole number $\ell$ for $\Lambda=0.003$.}\label{fig_l}
\end{figure}

In Fig.\ref{fig_l} , the QNMs are plotted as a function of multipole index $\ell$ for $\Lambda = 0.003$, charge $q = 0.4$ and overtone number $n$ = 0. It is found that ${\rm{Re}}~ \omega$ increases linearly with $\ell$, while magnitude of ${\rm{Im}}~ \omega$ initially increases rapidly with $\ell$ and later on, it saturates.	
	
	\begin{figure}[h]
		\centering
		\begin{minipage}{.5\textwidth}
			\centering
			\includegraphics[height=5 cm,width= 7 cm]{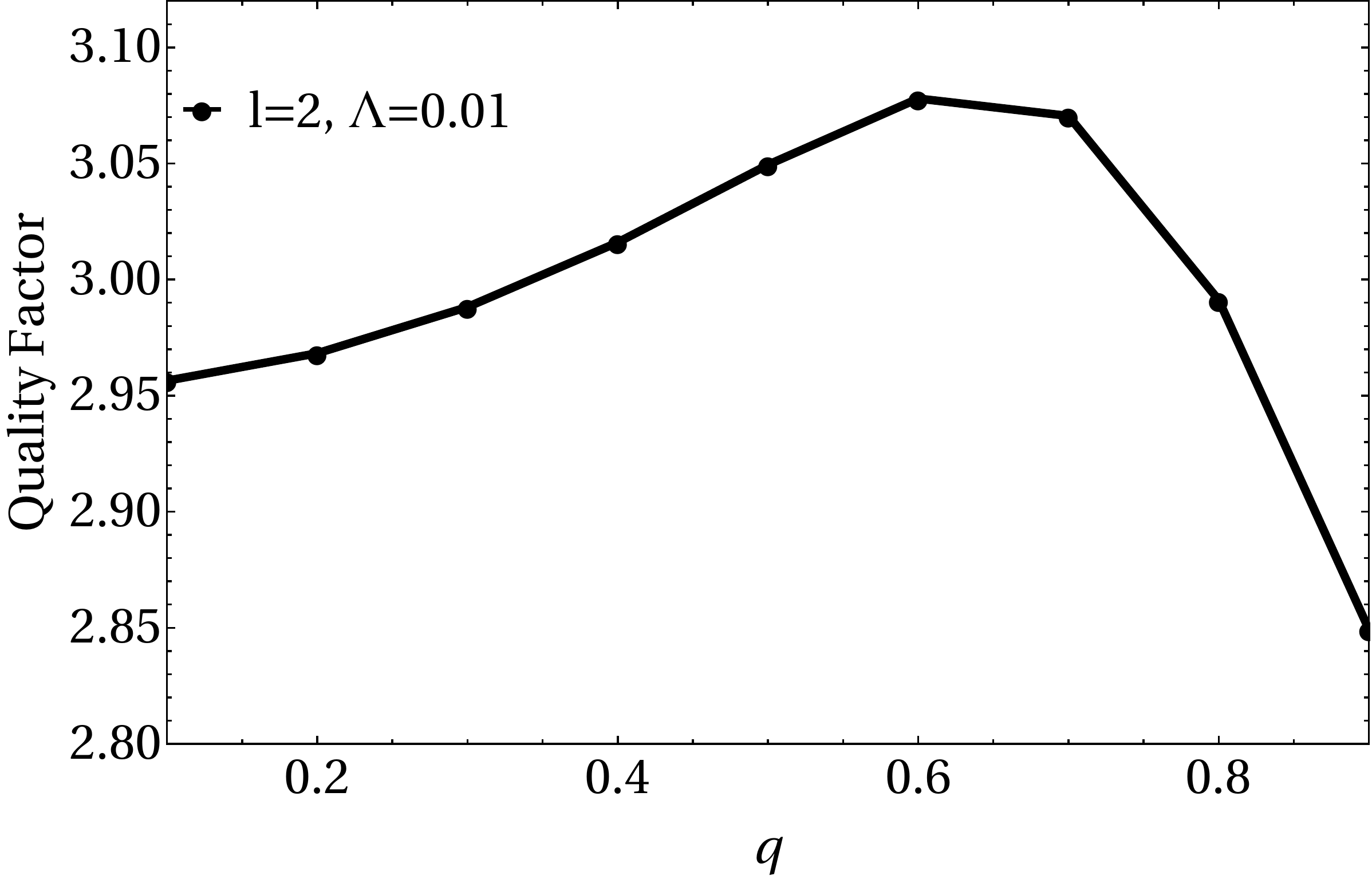}
		\end{minipage}%
		\begin{minipage}{.5\textwidth}
			\centering
			\includegraphics[height=5 cm,width= 7 cm]{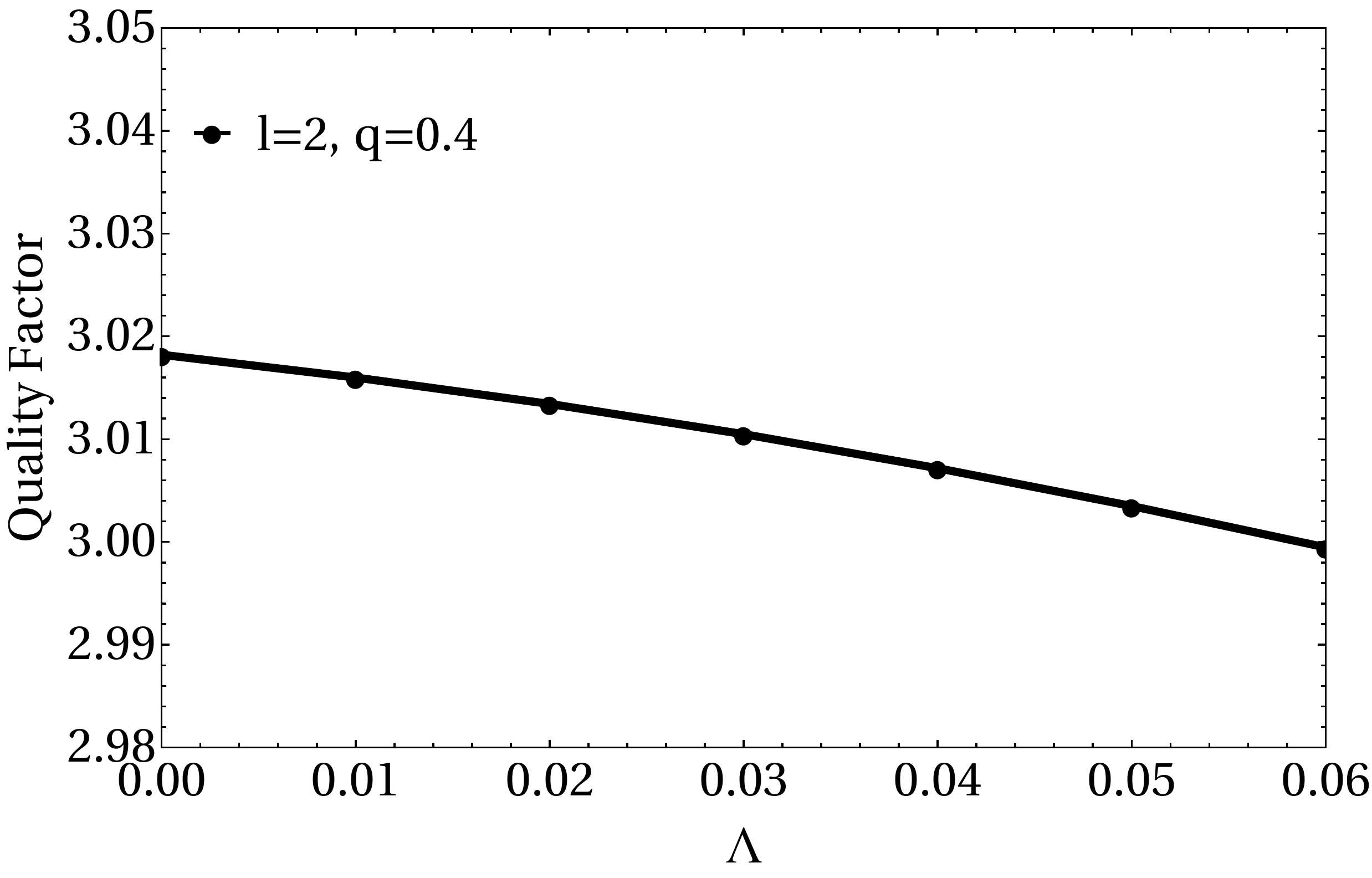}
		\end{minipage}
		\caption{Q-Factor vs magnetic charge $q$ and cosmological constant $\Lambda$.}\label{figure4}
	\end{figure}
Utilising the master eqn. (\ref{qnmeqn}), we have determined the QNMs for different set of parameters in this work. One can define the quality factor (Q.F.) to look at the strength of the field oscillation over damping as follows: ${\rm{Q.F.}} = \frac{{\rm{Re}}(\omega)}{2|{\rm{Im}}(\omega)|}$. It is well known that the quality factor is essentially a dimensionless parameter that describes how underdamped an oscillator is. In Fig.\ref{figure4} , we have plotted the Q.F. versus the charge $q$ and cosmological constant $\Lambda$. It is easy to check that field oscillation initially increases and finally decreases with $q$ for $\ell=2$ and $n=0$ but it decreases throughout the variation of $\Lambda$. This implies that the BdS black hole system becomes over-damped with the increase of cosmological constant. 
\begin{figure}[h]
		\centering
		\begin{minipage}{.5\textwidth}
			\centering
			\includegraphics[height=5 cm,width= 7 cm]{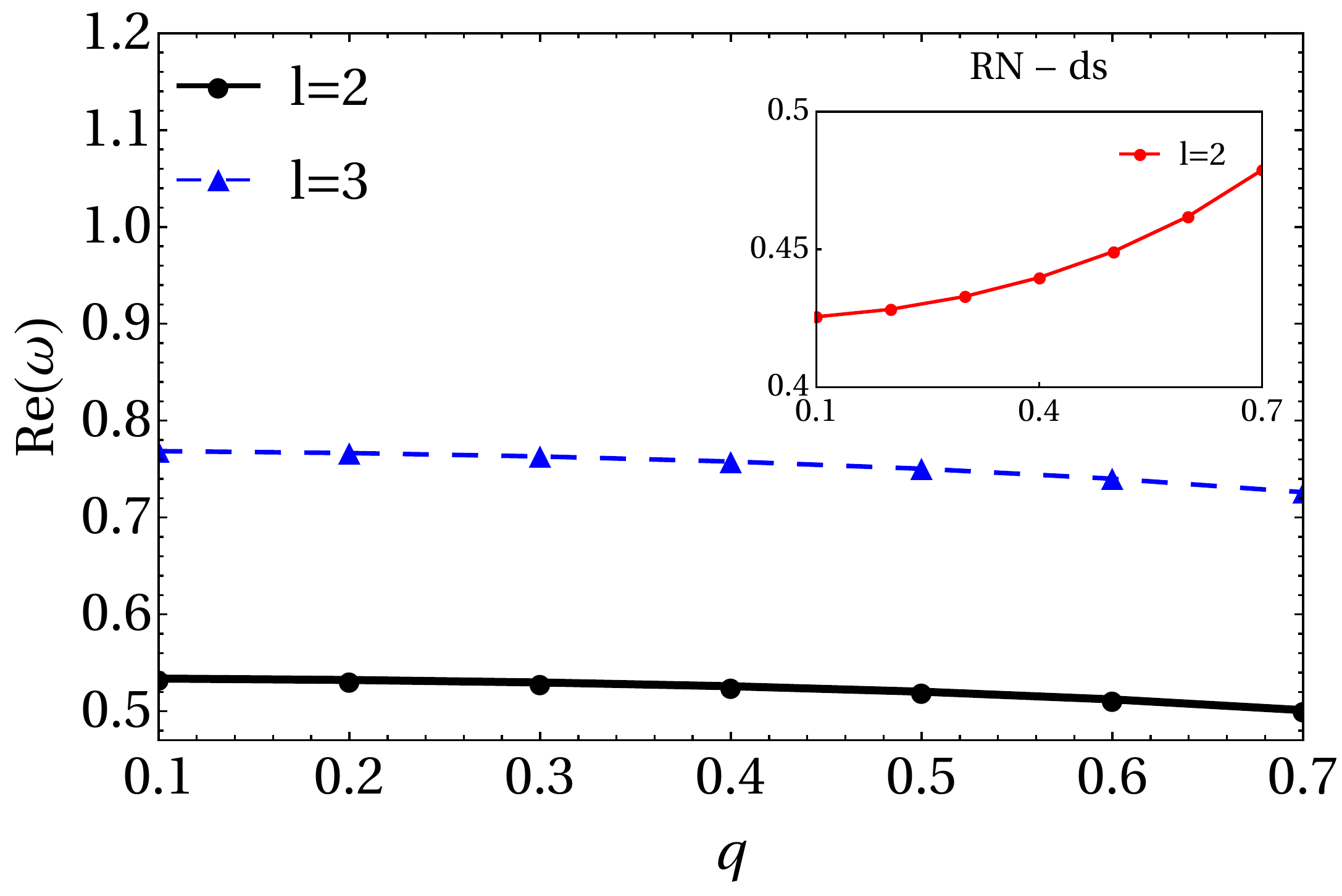}
		\end{minipage}%
		\begin{minipage}{.5\textwidth}
			\centering
			\includegraphics[height=4.7 cm,width= 7 cm]{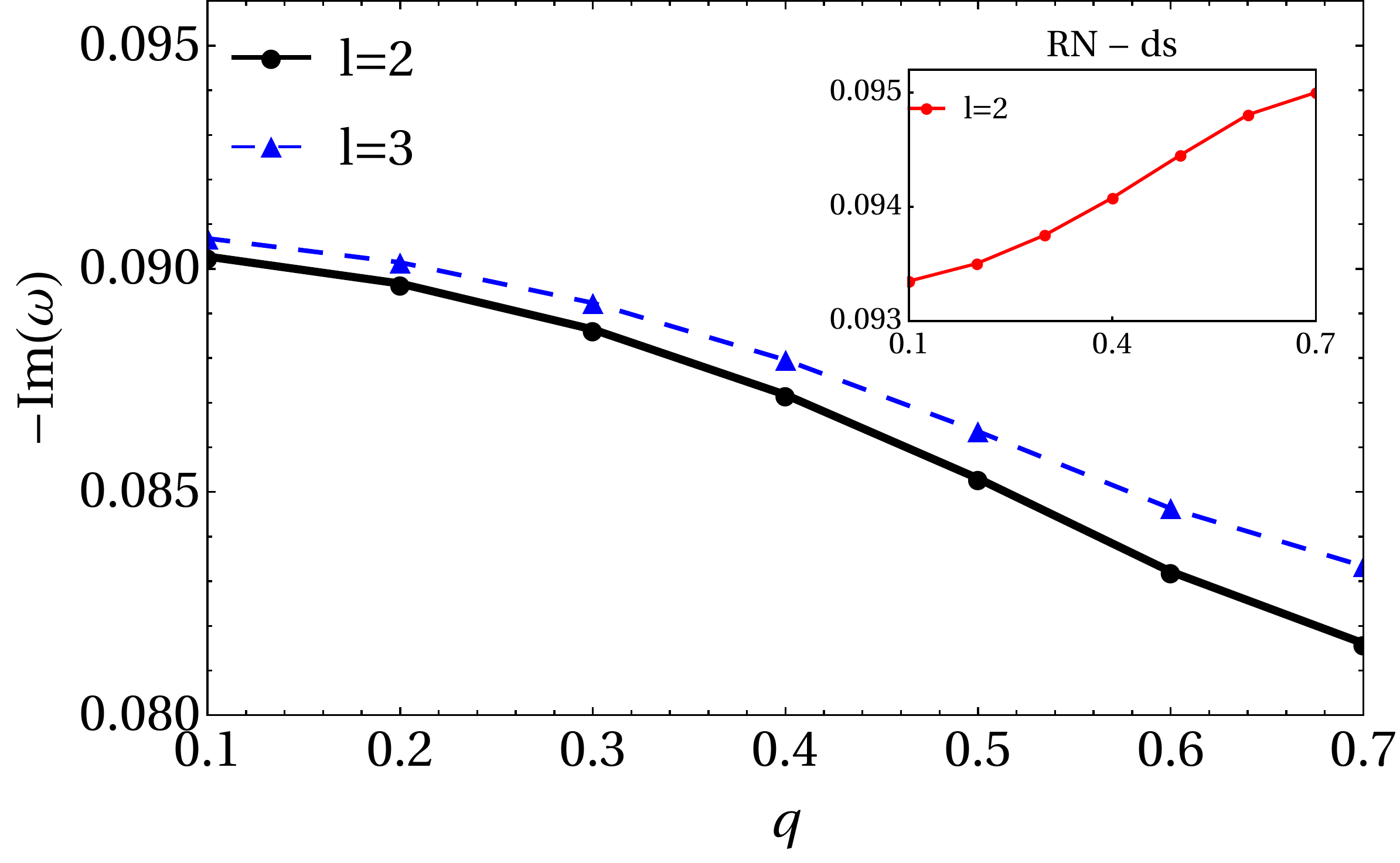}
		\end{minipage}
		\caption{Variation of Re($\omega$) and -Im($\omega$) vs. charge $q$ for a fixed value of $\Lambda =0.01$ for BdS black holes. The inset shows the same plott for RN-dS with the same values of the parameter.}\label{figure5}
	\end{figure}
Next, we plot the variation of QN frequencies with respect to charge $q$ and cosmological constant $\Lambda$ for different multipole numbers ($\ell$). Fig.\ref{figure5} specifically suggests the nature of QNMs as a function of $q$. Here for BdS, Re $\omega$ decreases constantly with $q$ but for RN-dS black hole (shown at the inset of the plot), it increases rapidly for the same parameter space. For Bds black hole, -Im $\omega$ declines abruptly with increasing $q$. On the contrary, it increases for RN-dS, which implies with smaller charge, BdS black hole is more stable than RN-dS. Whereas, Fig.\ref{figure6} demonstrates linear decrement in both real and imaginary part of QNMs with increasing $\Lambda$ for all sets of multipole numbers $\ell$=$1$ and $2$. Nature of the response from RN-dS black hole is same as BdS but its oscillation frequency is much smaller than BdS black hole keeping the nature of damping with respect to the parameters the same. Finally, in Table[\ref{table1}], we have listed the numerical values of QN frequencies which are obtained using sixth order WKB approach for the parameter $\Lambda=0.007$ and $q=0.57$. As it is well known that WKB method is accurate for $n<\ell$, we have tabulated the QN frequencies considering this condition. Data of Table[\ref{table1}] shows as $\ell$ increases both Re $\omega$ and -Im $\omega$ increase for a fixed overtone number ($n$). Another aspect of listed QNMs is that real oscillation frequency and imaginary part of the frequency representing damping are decreasing and increasing respectively with increasing overtone number $n$ for fixed $\ell$ values. This behaviour of QN frequencies with $n$ and $\ell$ is same amongst all different types of perturbations: electromagnetic, gravitation, massless and massive scalar perturbations \cite{wadbor}. 
	\begin{figure}[h]
		\centering
		\begin{minipage}{.5\textwidth}
			\centering
			\includegraphics[height=4.8 cm,width=6.8cm]{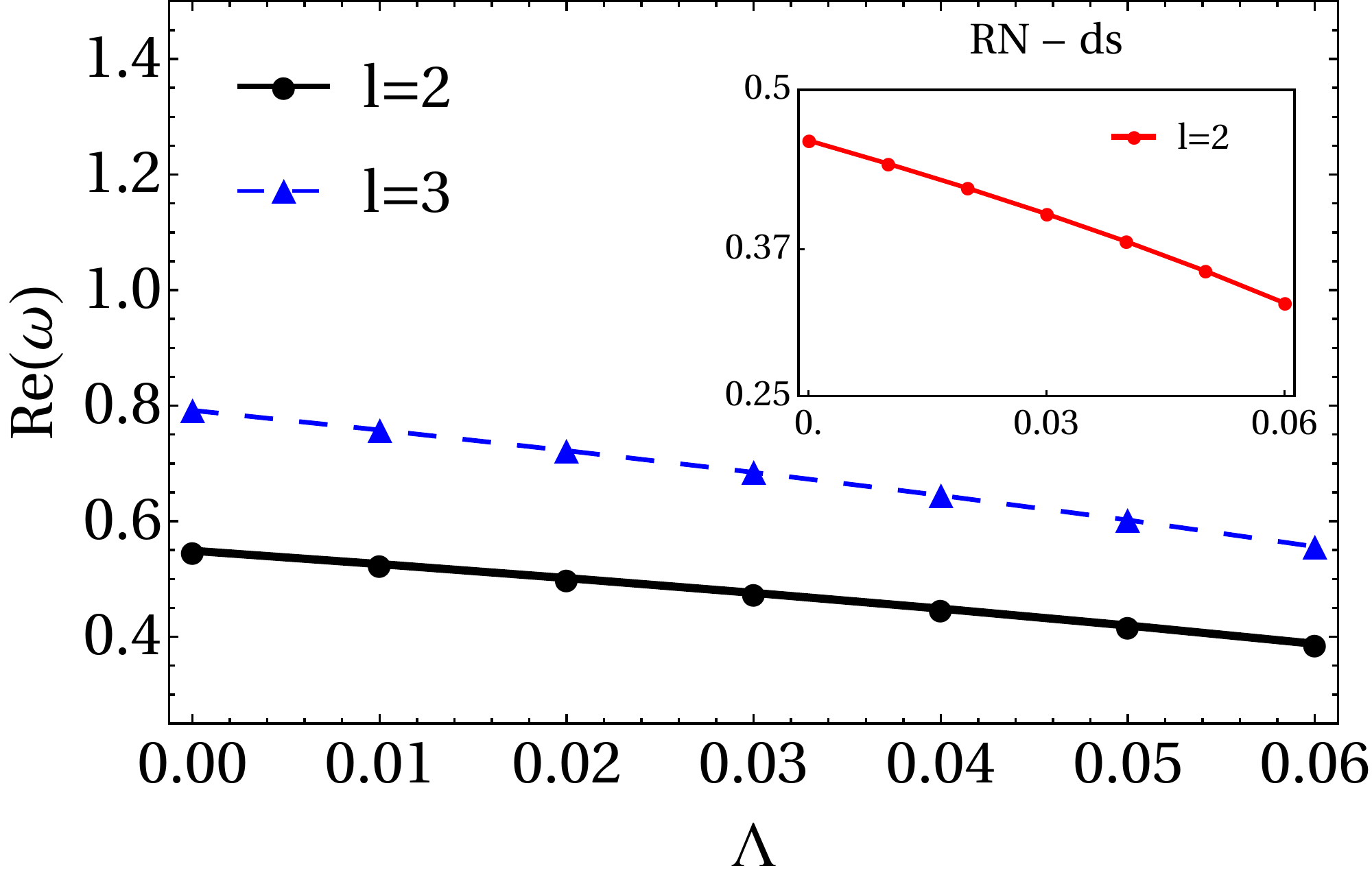}
		\end{minipage}%
		\begin{minipage}{.5\textwidth}
			\centering
			\includegraphics[height=5 cm,width=7cm]{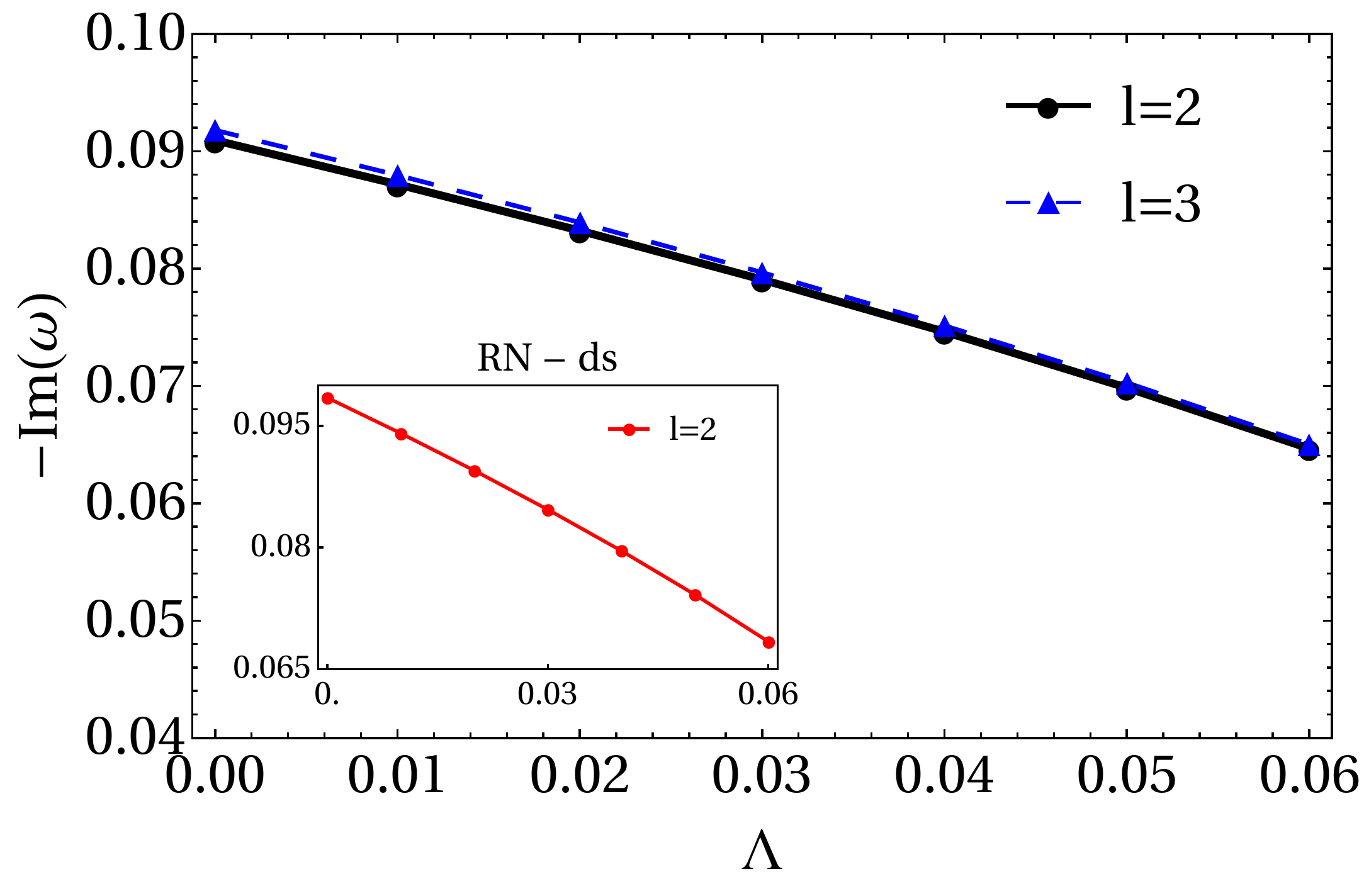}
		\end{minipage}
		\caption{Variation of Re($\omega$) and -Im($\omega$) vs $\Lambda$ with a fixed value of magnetic charge $q=0.4$}\label{figure6}
	\end{figure}
	\begin{table}[h]
		\begin{center}
			\begin{tabular}{|c|c|c|}
				\hline
				\hline
				Multipole number & Overtone & QN frequencies using 6th order WKB \\
				\hline
				$\ell=2$& n$=0$ &$0.521390 - 0.084842$i \\
				& n$=1$ &$0.507453 -0.256951$i\\
				\hline
				&n$=0$&$0.753028 - 0.086184$i\\
				
				$\ell=3$&n$=1$&$0.742394 - 0.259872$i\\
				
				&n$=2$&$0.721758 - 0.437509$i\\
				\hline
				&n$=0$&$0.978881 - 0.086878$i\\
				
				&n$=1$&$0.970469 - 0.261448$i\\
				
				$\ell=4$&n$=2$&$0.953945 -0.438453$i\\
				
				&n$=3$&$0.929943 -0.619473$i\\
				\hline
				&n$=0$&$1.202785 - 0.087246$i\\
				
				&n$=1$&$1.195857 - 0.262286$i\\
				
				$\ell=5$&n$=2$&$1.182170 - 0.438968$i\\
				
				&n$=3$&$1.162055 - 0.618374$i\\
				
				&n$=4$&$1.136057  - 0.801537$i\\
				\hline 
			\hline
			\end{tabular}
		\end{center}
		\caption{Electromagnetic QN frequencies for the Bardeen de-Sitter black hole spacetime as a function of $\ell$ and $n$ for $q=0.57$ and $\Lambda=0.007$. }\label{table1}
	\end{table}
	It is worth mentioning here that by computing inverse of the instability timescale which is associated with the geodesic motion, it is possible to show that in the eikonal limit, parameters of the circular null geodesics can determine the QNMs of black holes in any dimensions \cite{cbwz}. This is a very important and strong result since the parameters of null geodesics can throw some light on the stability of a black hole. It has also been shown to be independent of the field equations. The only assumption which went into the consideration of the authors of \cite{cbwz}, is the fact that the black hole spacetime is static, spherically symmetric and asymptotically flat. However as a non-trivial example, they have discussed non-asymptotically flat near extremal Schwarzschild de Sitter black hole space time in this context.  Therefore, the same analysis can be applied for BdS black holes in the limit of near extremal  regime (Nariai or cold black holes) where either the black hole horizon and the cosmological horizon coincides or the inner and outer horizon merges. 

\section{Gravitational perturbations and QNMs of the BdS black hole}

It has to be mentioned here that generally there are two different categories of perturbations of black holes that are considered within the regime of general theory of relativity. In the first method, one adds a test field in a black hole background and the system is studied by solving the dynamical equation for the particular test field in the background of the black hole. The second one is to perturb the metric itself and in order to find the evolution equations, one linearises the Einstein's equations. This is the gravitational perturbation and is the most important one amongst all types of perturbations since the gravitational radiation is much stronger than strength of any external fields decaying near the black hole. It is also important because the metric perturbations gives us tools to study  about the gravitational stability of a black hole. The investigation of black hole perturbations was first carried out by Regge and Wheeler \cite{rg} for the odd parity type of the spherical harmonics and was extended to the even parity type by Zerilli \cite{zr}. A brief discussion about the calculations involved in gravitational perturbation is given in the appendix. 
\begin{figure}[h]
 \centering
\begin{minipage}{.5\textwidth}
  \centering
\includegraphics[height=6 cm,width= 7.5 cm]{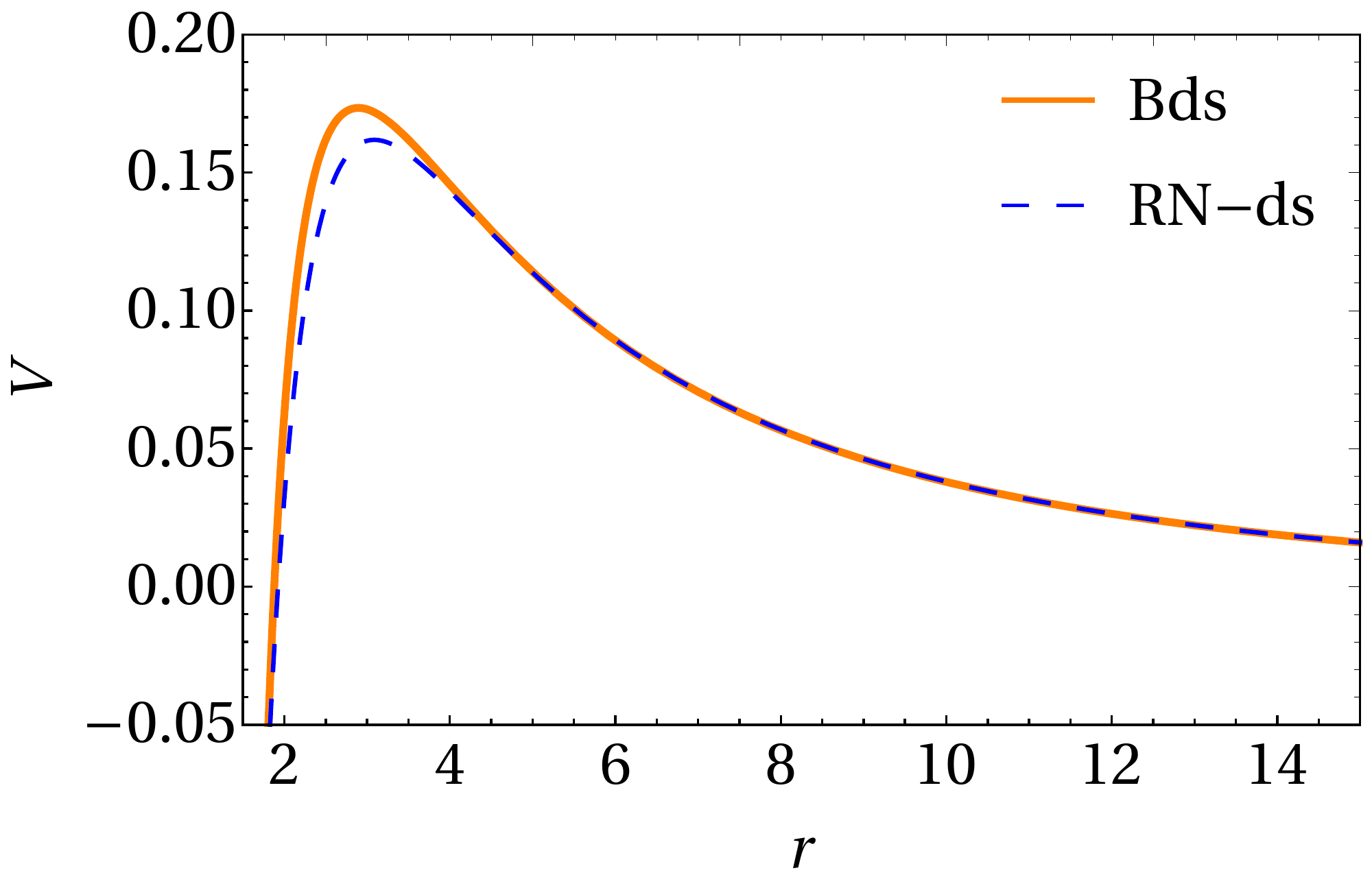}
\end{minipage}
\caption{Variation of gravitational potential $V$ in  Regge-Wheeler gauge with $r$ for $\ell=2$, $\Lambda=0.003$ and $q=0.40$.}\label{figure7}
\end{figure}
The form of the potential due to gravitational perturbation is given by 
\begin{equation}
 V(r)=f \left[\frac {\ell(\ell+1)+r \left(rf'\right)'+2(f-1)+2 r^2 (2\mathcal{L}+\Lambda)} {r^2}\right]
\end{equation}
\begin{figure}[h]
\centering
\begin{minipage}{.5\textwidth}
 \centering
  \includegraphics[height=5 cm,width= 7 cm]{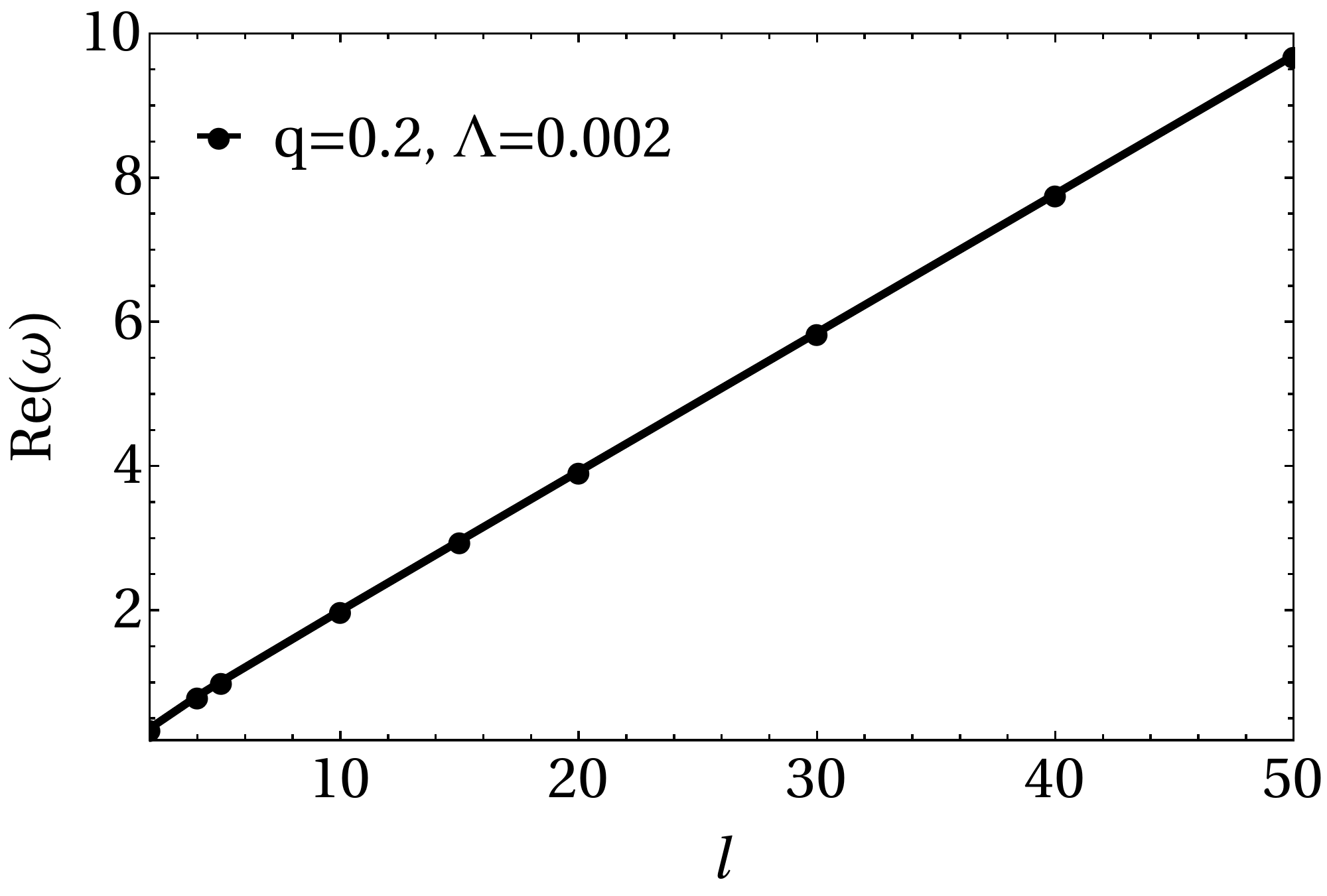}
\end{minipage}%
\begin{minipage}{.5\textwidth}
  \centering
   \includegraphics[height=5 cm,width= 7 cm]{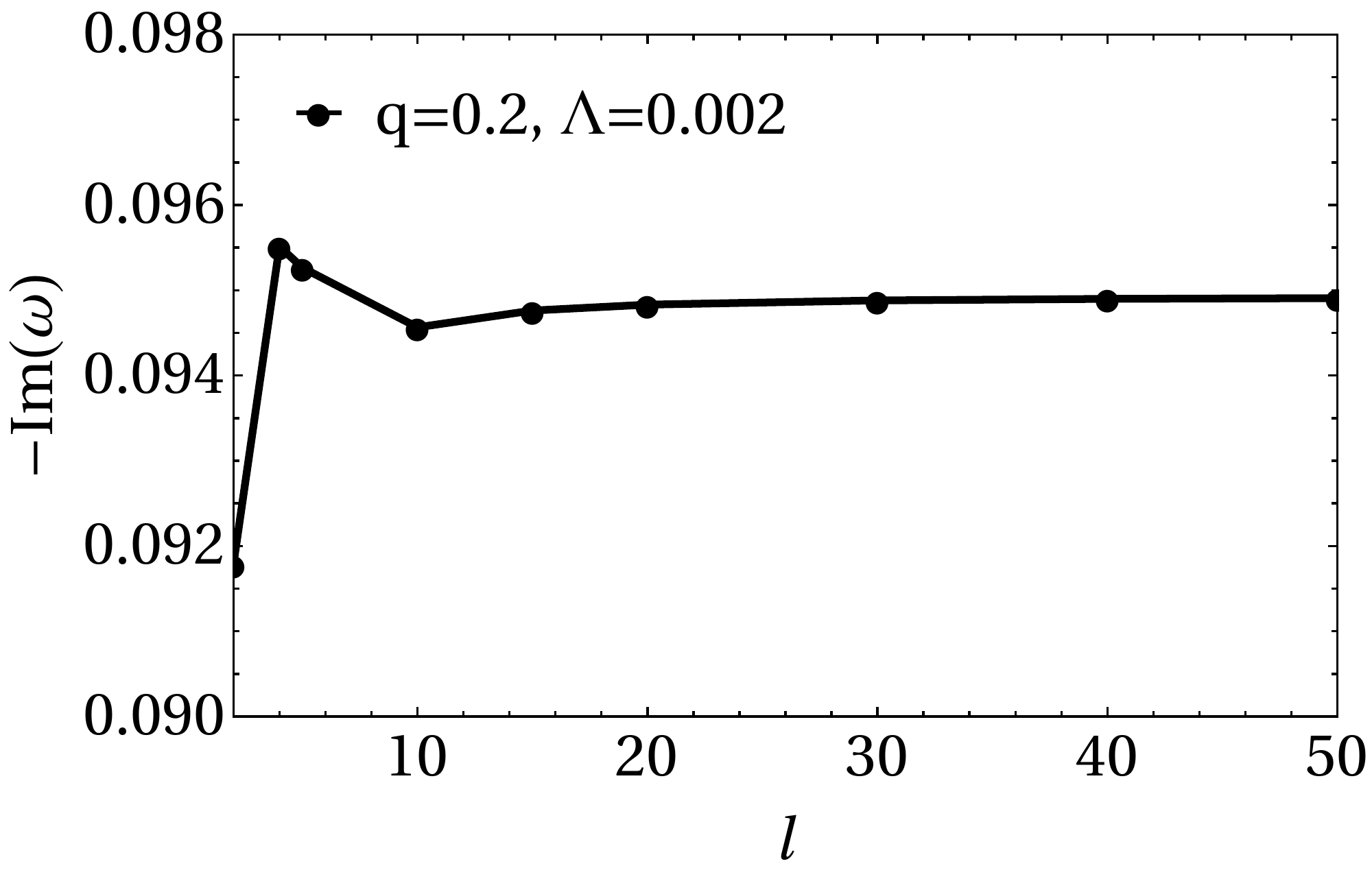}
\end{minipage}
       \caption{Variation of Re $\omega$ and -Im $\omega$  with multipole number $\ell$ for $\Lambda=0.002$ and $n=0$.}\label{figure9}
\end{figure}
where, $\mathcal{L}$ denotes lagrangian of the field and $\Lambda$ is cosmological constant.\newline
 In Fig.\ref{figure7} , we have shown the nature of the effective potential $V$ for BdS black hole with fixed values of the parameters $\Lambda=0.003$, $\ell=2$ and $q=0.2$. For comparison between black holes in linear and nonlinear electromagnetic field, we have studied RN-dS and BdS black holes respectively. As before, the value of the mass of the black hole is taken as unity throughout this paper. Like the electromagnetic, here $V$ is also positive, finite which increases its height with increasing $\ell$ and almost same spatial extent for both BdS and RN-dS black hole. But height of the BdS potential is always larger than RN-dS for which total absorption cross-section of BdS black hole is always less than its counter part in linear electromagnetic field. With $\ell = 0, $ $V$ has more than one extremum which prevents us to apply the WKB approach. Therefore, like electromagnetic perturbation, here also we will be considering $\ell \neq 0$ modes.
In Fig.(\ref{figure9}), the QNMs are plotted as a function of multipole index $\ell$ for $\Lambda=0.002$, magnetic charge $q=0.2$ and overtone number  $n=0$. It is found that Re $\omega$ increases linearly with $\ell$ while magnitude of Im $\omega$ initially increases rapidly with $\ell$ and later on, it saturates and becomes almost a constant. Although the behaviour of the frequencies remains similar to those of electromagnetic ones as we vary the multipole index, the rapidity with which the imaginary parts of the frequency change with $\ell$ in case of the gravitational perturbation is much higher than that of the electromagnetic case. 
\begin{figure}[h]
\centering
\begin{minipage}{.5\textwidth}
 \centering
  \includegraphics[height=5 cm,width= 7 cm]{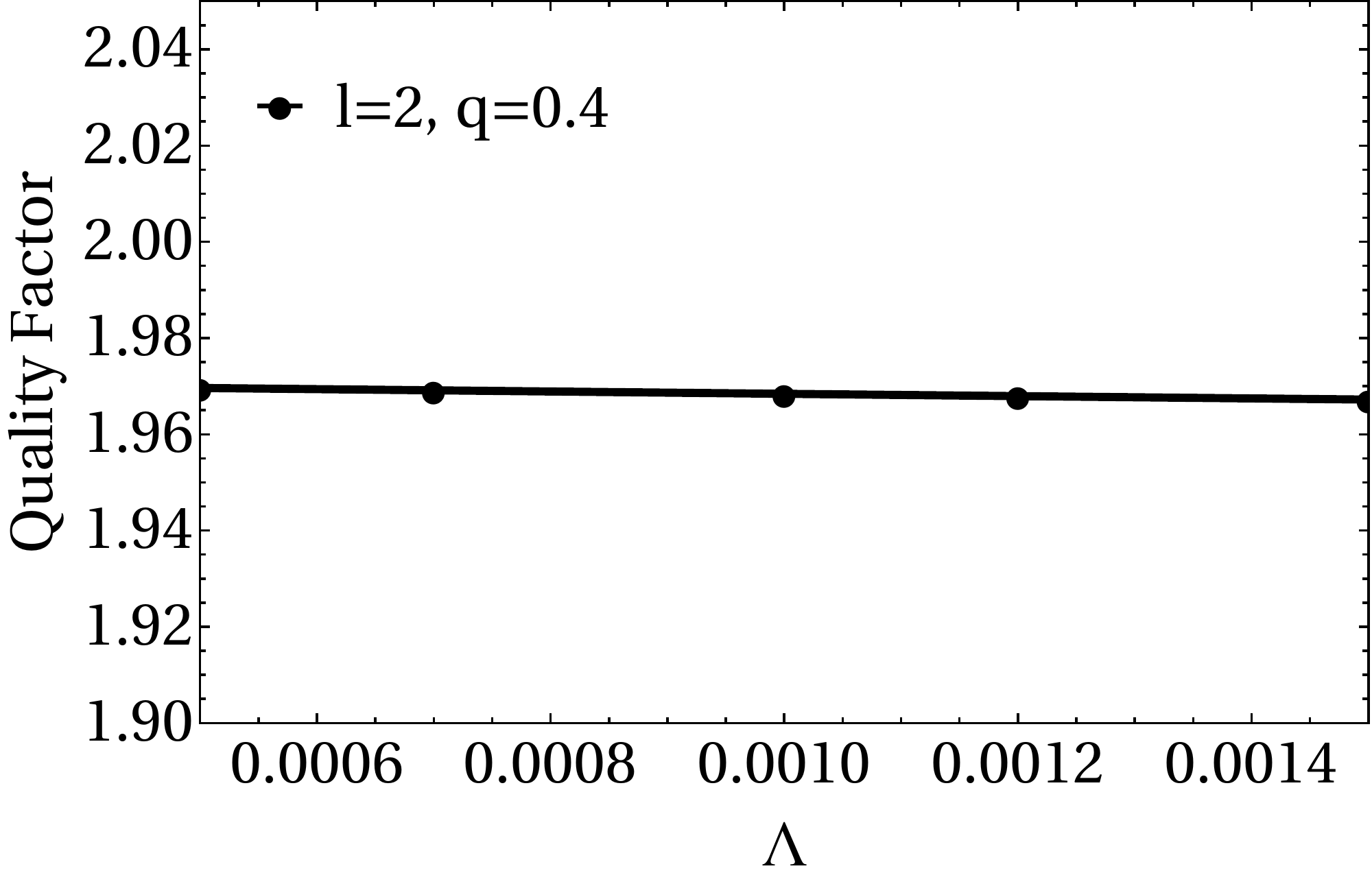}
\end{minipage}%
\begin{minipage}{.5\textwidth}
  \centering
  \includegraphics[height=5 cm,width= 7 cm]{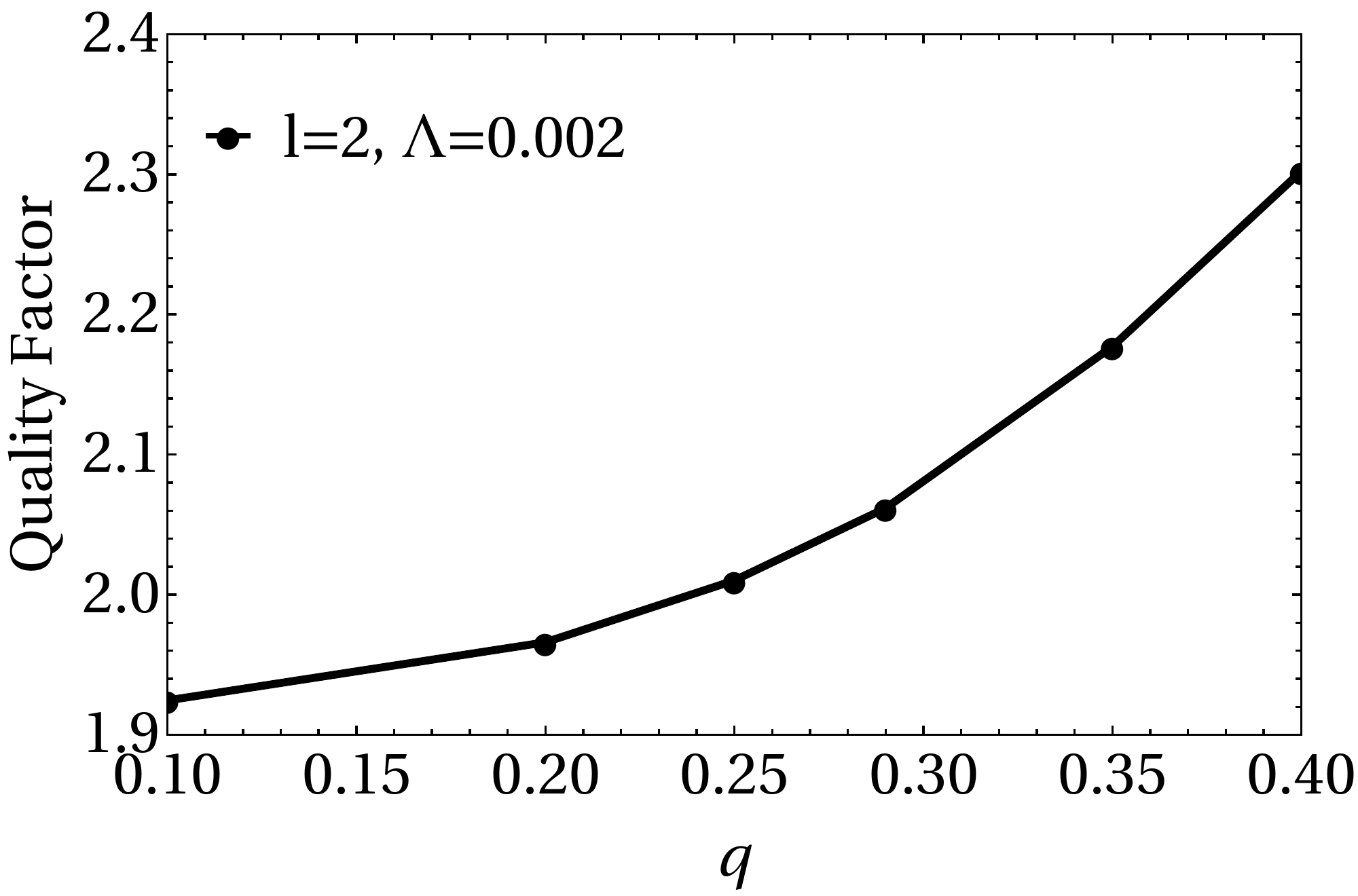}
\end{minipage}
\caption{Q-factor vs parameters $\Lambda$ and $q$}\label{figure10}
\end{figure}
\begin{figure}[h]
  \centering
 \begin{minipage}{.5\textwidth}
  \centering
 \includegraphics[height=5 cm,width= 7 cm]{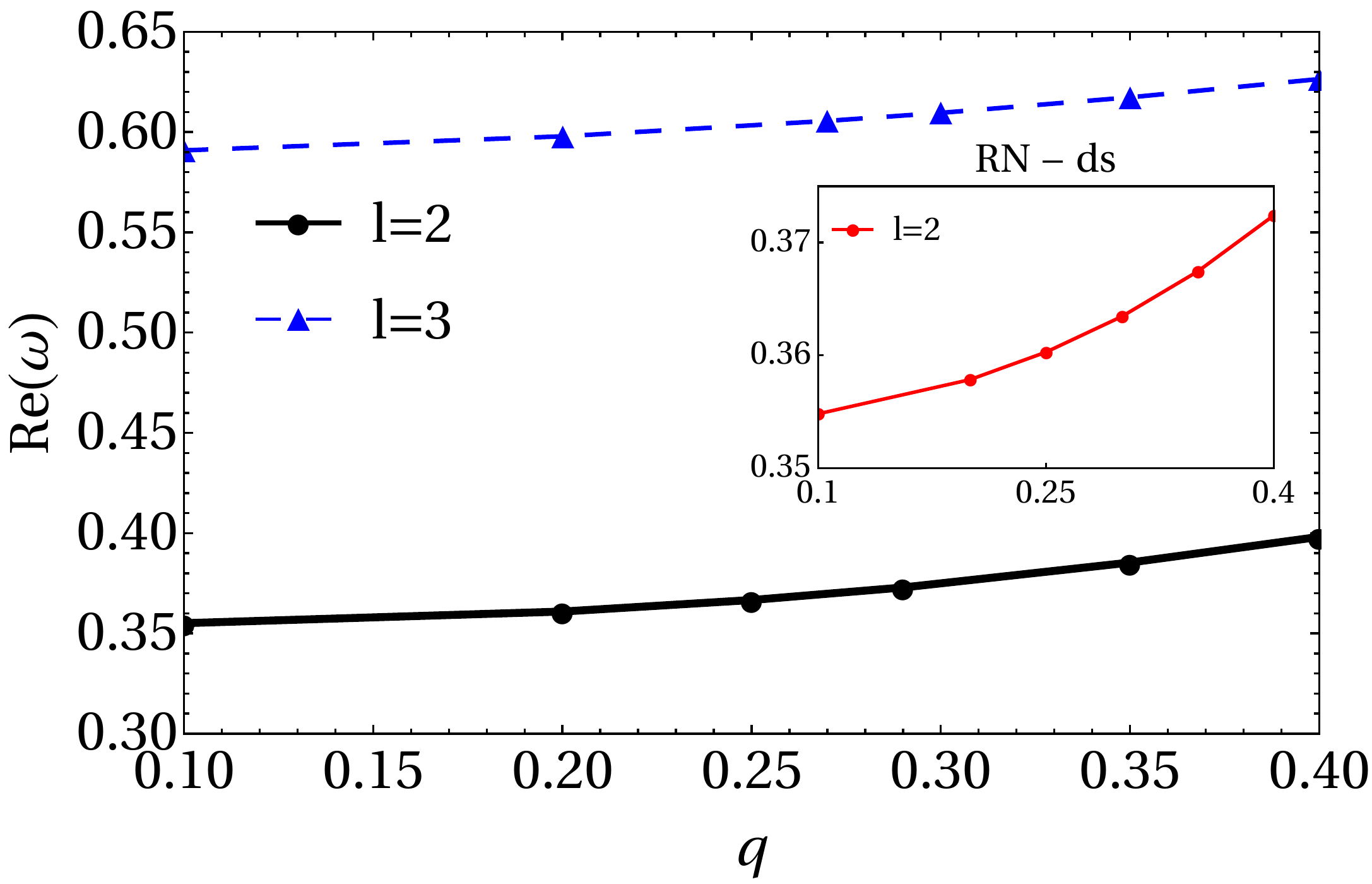}
 \end{minipage}%
  \begin{minipage}{.5\textwidth}
 \centering
 \includegraphics[height=5 cm,width= 7 cm]{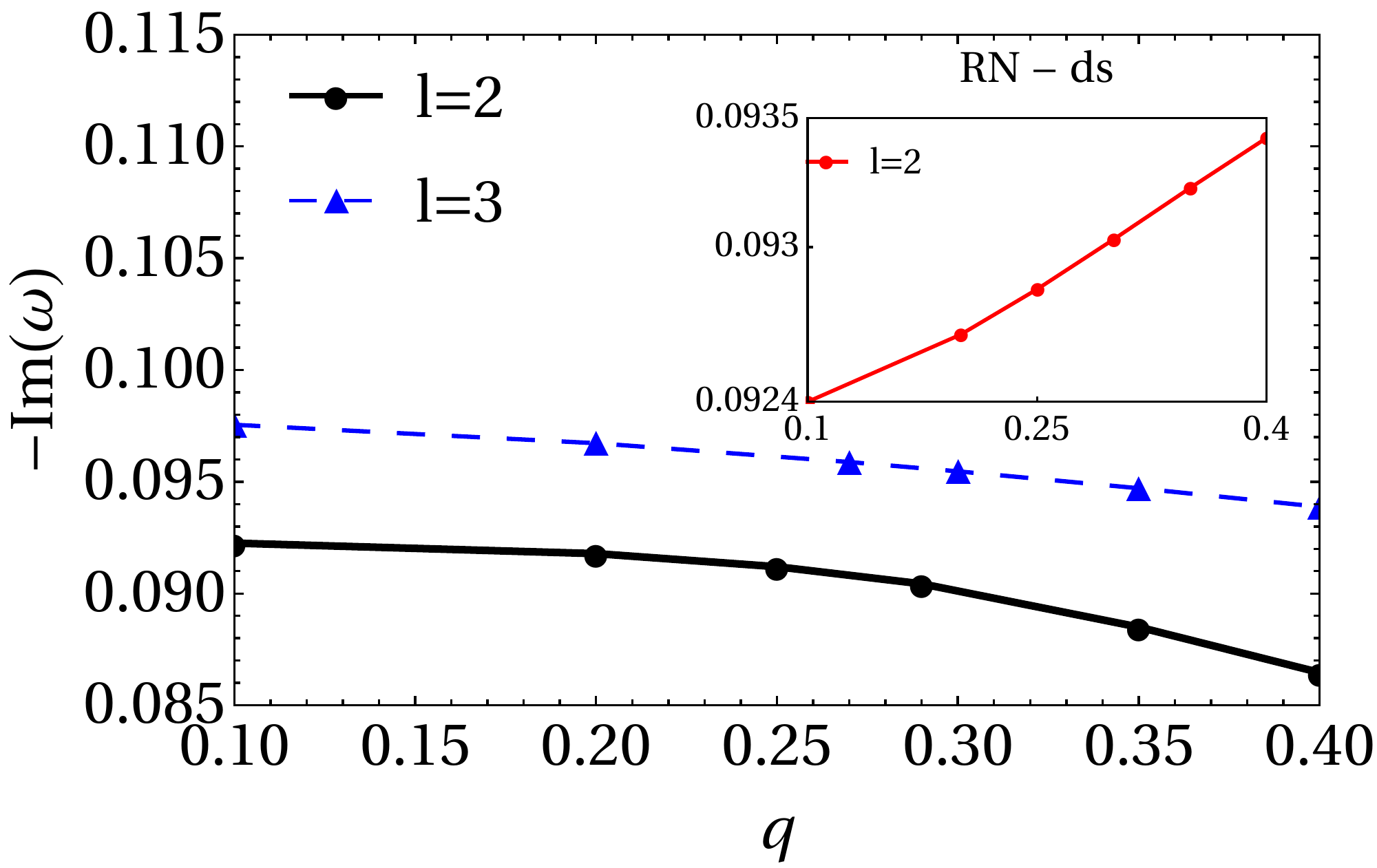}
 \end{minipage}
 \caption{Variation of Re($\omega$) and -Im($\omega$) vs $q$, for $ \Lambda =0.002 $.}\label{figure11}
 \end{figure}To understand the strength of the gravitational perturbation, we plot the quality factor (Q.F.) versus cosmological constant $\Lambda$ and magnetic charge $q$ in Fig.\ref{figure10} . It is clear from the plot that field oscillation is almost same in a with variation of $ \Lambda $ making a significant difference with it's electromagnetic counterpart for $ q=0.4$. At the same time, variation of Q.F. with magnetic charge($q$) shows us a nonlinear increment behaviour as we increase $q$ for a fixed value of $\Lambda=0.002$. Next, we plot QN frequencies of the BdS black hole for gravitational perturbation vs magnetic charge parameter ($q$) and cosmological constant ($\Lambda$) with different $\ell$ values. Fig.\ref{figure11} precisely suggests the nature of QNMs as a function of $q$. For BdS, Re $\omega$ follows non-linear relation with $q$ keeping similarity with its linear electromagnetic counterpart(RN-dS). However, the magnitude of Im $\omega$ falls with increasing magnetic charge $q$. It clearly shows difference in nature of dependence between BdS and RN-dS black hole. For the same set of parameters, with increasing magnetic charge($q$) BdS black hole becomes unstable than RN-dS black hole. 
\begin{figure}[H]
  \centering
 \begin{minipage}{.5\textwidth}
  \centering
 \includegraphics[height=5 cm,width= 7 cm]{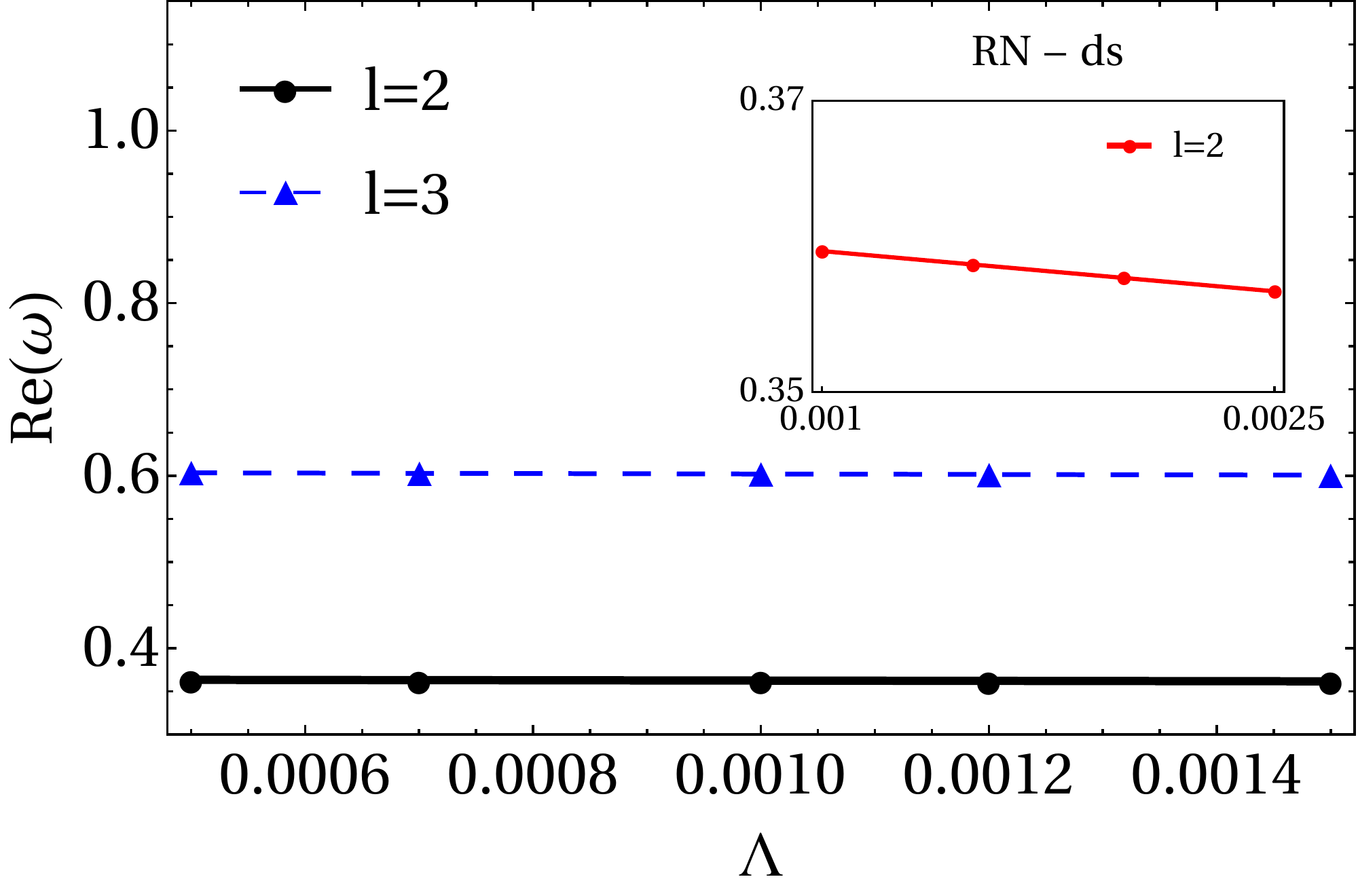}
 \end{minipage}%
  \begin{minipage}{.5\textwidth}
 \centering
 \includegraphics[height=5 cm,width= 7 cm]{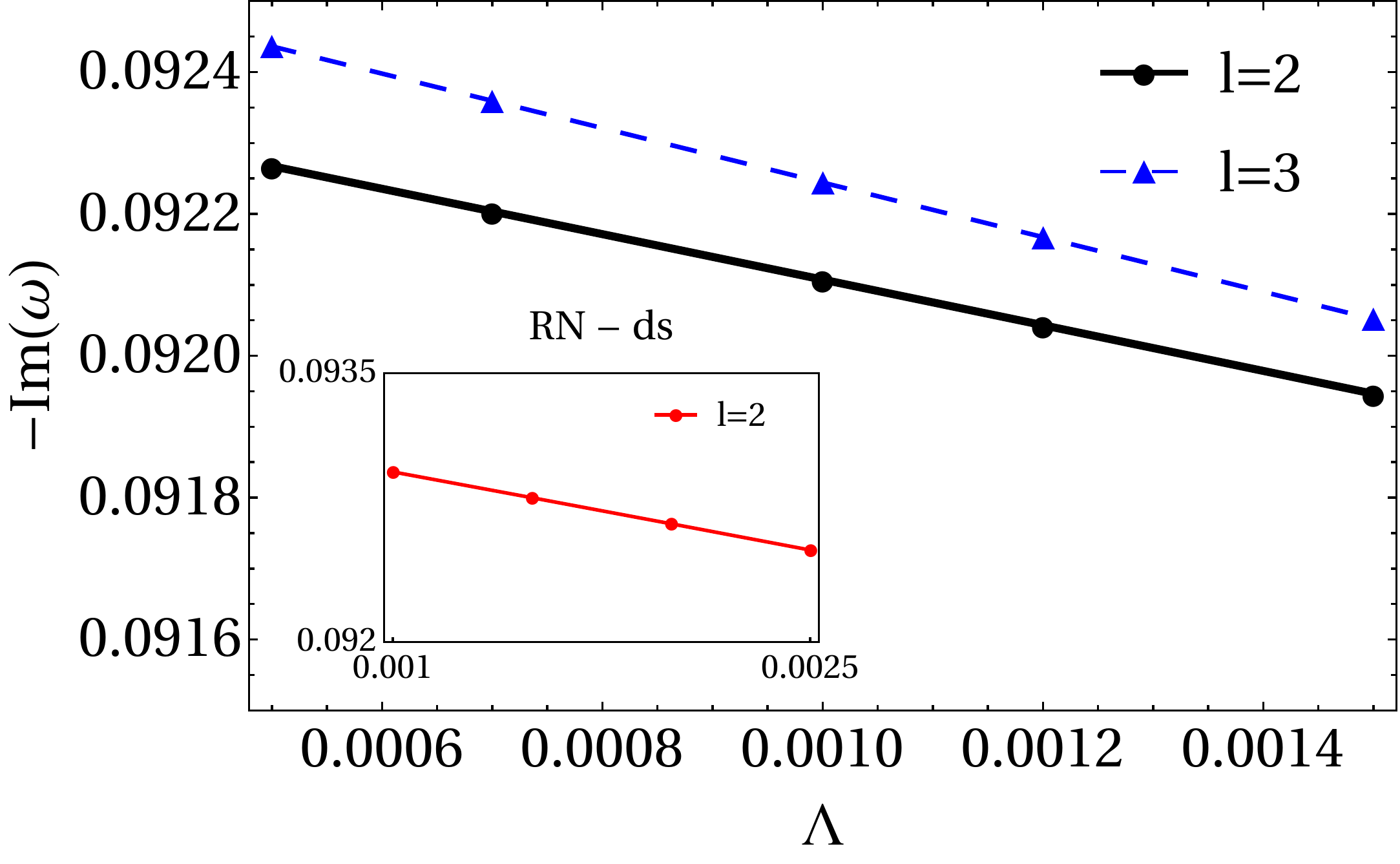}
 \end{minipage}
 \caption{Variation of Re($\omega$) and -Im($\omega$) vs $\Lambda$ for $q=0.2$.}\label{figure12}
\end{figure}
\begin{table}[h]
\begin{center}
\vspace{0.8 cm}
\begin{tabular}{|c|c|c|}
\hline
\hline
\textsf{Multipole number}& Overtone& QN frequencies using 6th order WKB\\
\hline
&n$=0$&$0.679040 - 0.089022$i\\

$\ell=3$ &n$=1$&$0.638172 - 0.274874$i\\
 &n$=2$&$0.357253 - 0.622339$i\\
\hline
 &n$=0$&$0.891988 - 0.089235$i\\
$\ell=4$ &n$=1$&$0.879436 - 0.271288 $i\\
 &n$=2$&$0.779455 - 0.472025$i\\
 &n$=3$&$0.363485 - 0.966099$i\\
\hline
 &n$=0$&$1.098212 - 0.092047$i\\
 &n$=1$&$1.057292 - 0.308499$i\\
$\ell=5$&n$=2$&$0.785209 - 0.807262$i\\
 &n$=3$&$0.487847 - 2.307081$i\\
 &n$=4$&$0.409467 - 4.815479$i\\
\hline
\hline
\end{tabular}
\end{center}
\caption{The list of gravitational QNMs in the Bardeen de-Sitter black hole space time as a function of $\ell$ and $n$ for $q=0.60$ and $\Lambda=0.003$.}\label{table2}
\end{table}
Fig.\ref{figure12} shows Re $\omega$ and $|$Im $\omega|$ decreases steadily for increasing $\Lambda$ for $q=0.4$ in both BdS and RN-dS black holes. Finally, In Table \ref{table2}, we have listed the numerical values of QN frequencies with corresponding parameters considering $n<\ell$. Like electromagnetic class (see Table \ref{table1}), tabulated QNMs indicate that as $\ell$ increases both real and modulus of imaginary $\omega$ increase for a fixed overtone number ($n$). Another feature of listed QNMs is that oscillation frequency and damping of perturbation are decreasing and increasing respectively when we increase $n$ keeping $\ell$ fixed. This behaviour is same irrespective of the class of perturbations.

\section{Dynamics of perturbation}
Our initial motivation was to study black hole stability under an external perturbation. C. V. Vishveshwara was the first person to realize
that we may observe a solitary black hole by observation of scattering of radiation from the black hole, provided the black hole left its fingerprint on the scattered wave \cite{visv}. Realising this, he started pelting the black hole with Gaussian wave packets and found that the black hole responds by ringing in a very unique decay mode: the lowest damped one of the black hole QNMs. Following this, here we will demonstrate a complete evolution picture of the BdS space time from a single master equation (see Eqn. \ref{grav_wave} discussed in the appendix). This is a wave equation with a schr\"{o}dinger like form. 
\begin{equation}
\frac{\partial ^2 \psi}{\partial t^2}-\frac{\partial ^2 \psi}{\partial r^2 _*}+V \psi=0\label{master_wave}
\end{equation}
We have used finite difference method to numerically integrate this wave equation(\ref{master_wave}). As a boundary condition we have Eqn. (\ref{boundary}) which describes asymptotic behaviour with pure ingoing and outgoing waves at $r_*\to \mp \infty$ respectively. We assume a solution of Eqn.(\ref{master_wave}) with oscillatory factor  $e^{-i\omega t}$ in time.
\begin{equation}
\lim_{r_*\to \pm \infty} \psi e^{\mp i \omega r_*} = 1\label{boundary}
\end{equation} 
To give the first external ``kick" in the field we use two Gaussian waves (\ref{int1}) and (\ref{int2}) as initial conditions for the second order differential equation(\ref{master_wave}):
\begin{equation}
\psi(r_*,0)=a_1 e ^{-\sigma_{1} (r_{*}-\mu)^2}\label{int1},
\end{equation}
\begin{equation}
\frac{\partial \psi(r_*,0)}{\partial t}=a_2 e ^{-\sigma_{2} (r_{*}-\mu)^2}\label{int2},
\end{equation}
\begin{figure}[H]
 \centering
\includegraphics[height=6 cm,width= 9 cm]{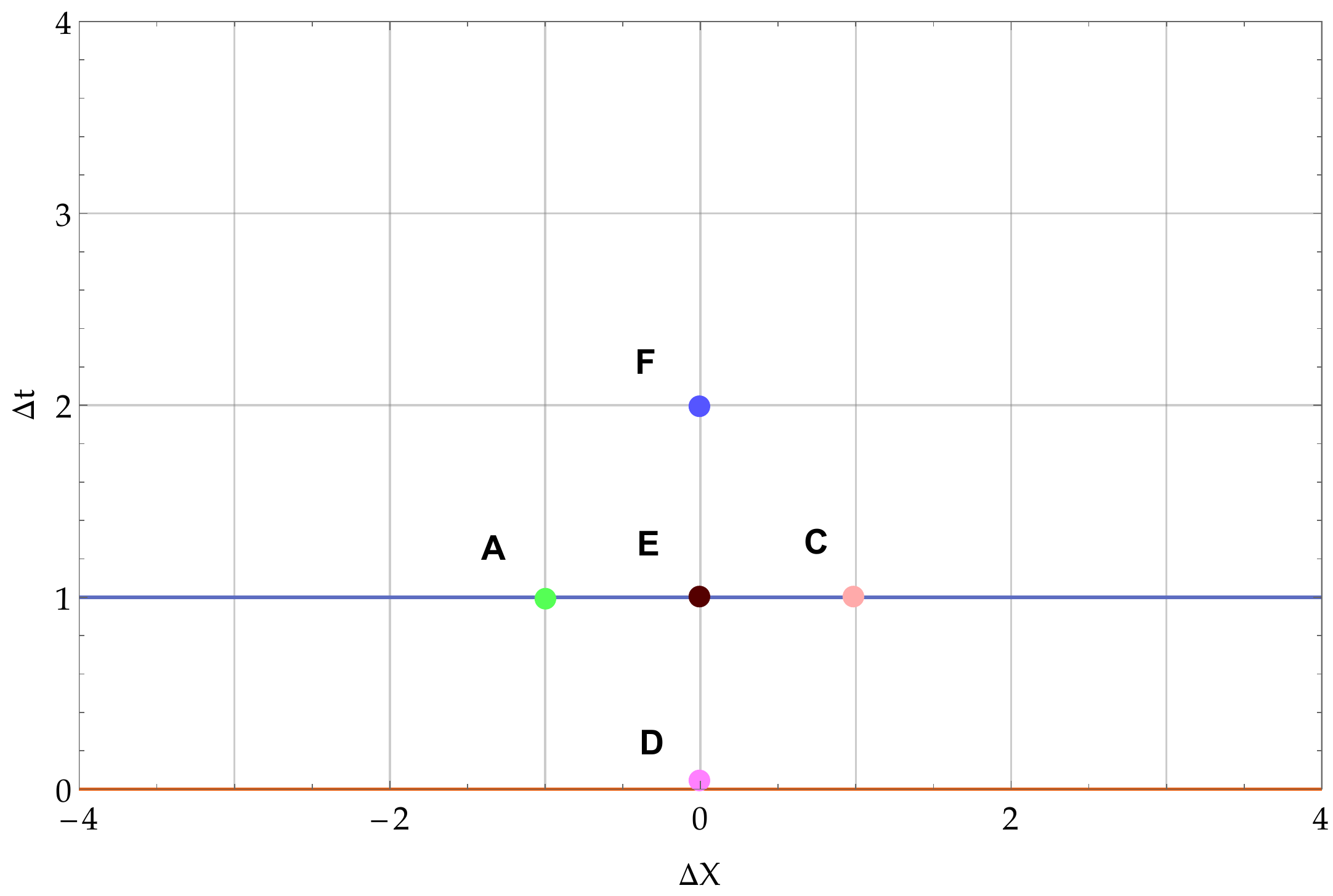}
\caption{$(x-t)$ plane for Integration scheme.}\label{grid_final}
\end{figure}
 \begin{figure}[h]
 \begin{minipage}{.5\textwidth}
 \includegraphics[height=5 cm,width= 7 cm]{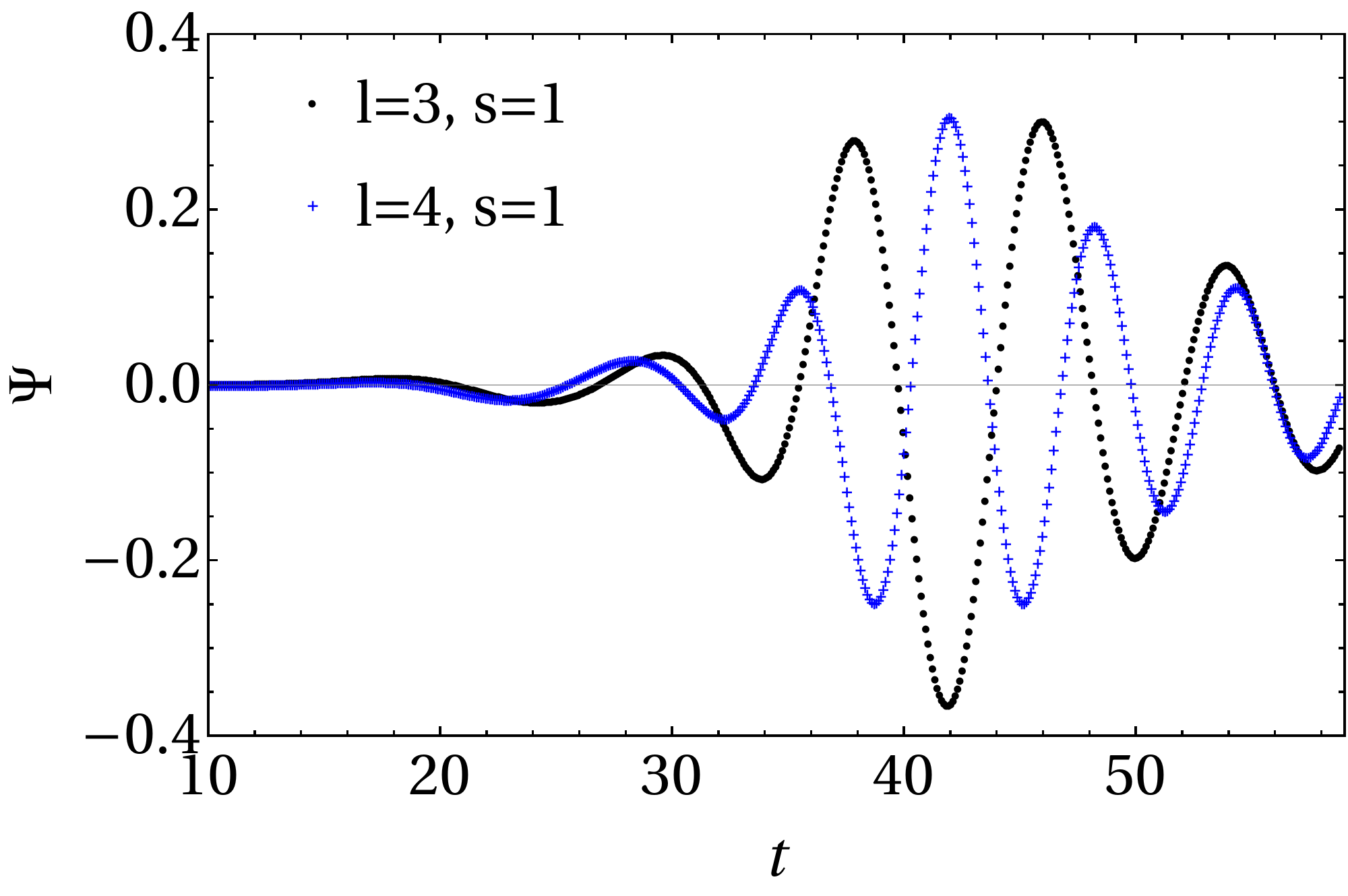}
 \end{minipage}%
  \begin{minipage}{.5\textwidth}
 \includegraphics[height=5 cm,width= 7 cm]{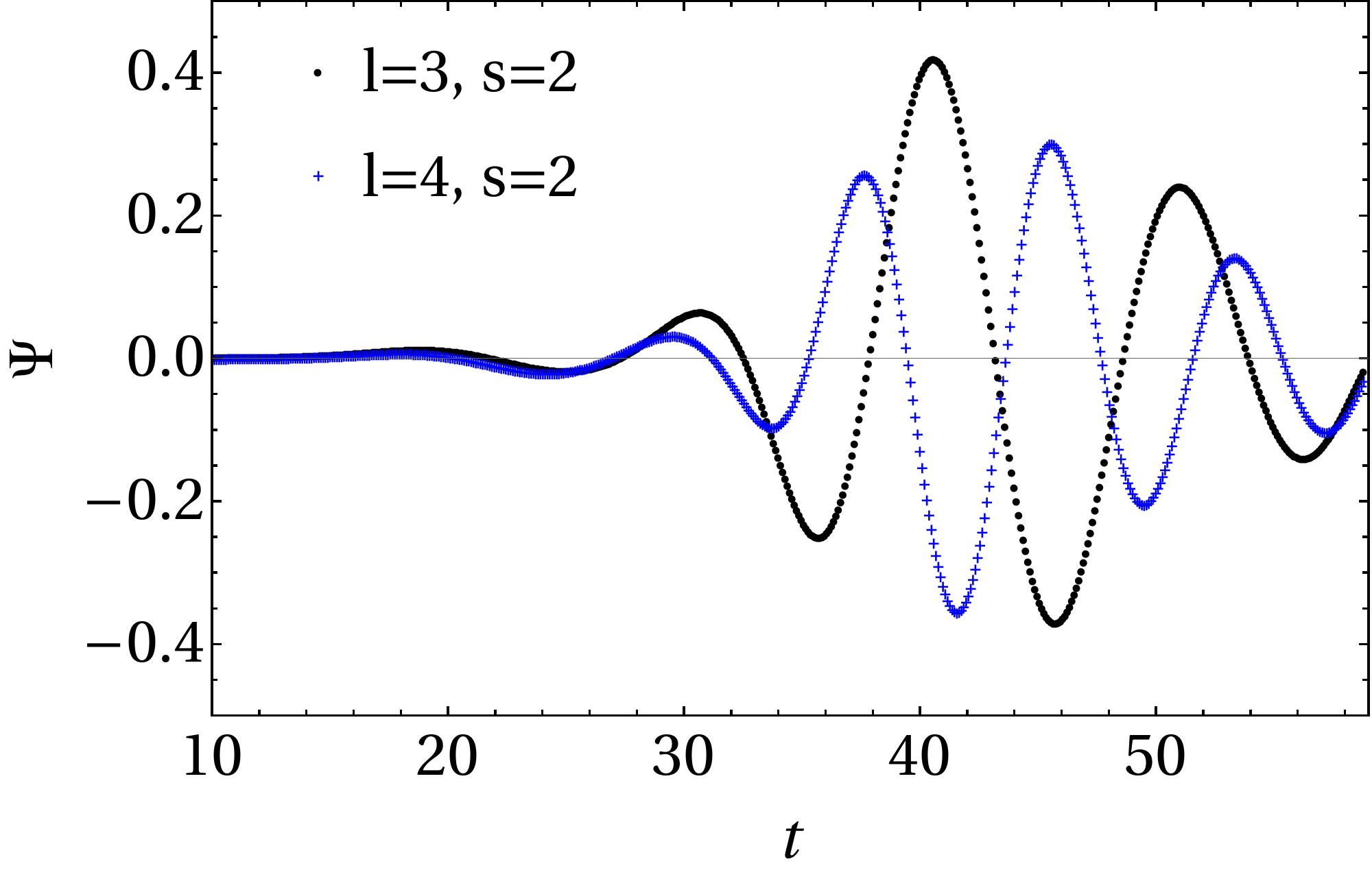}
 \end{minipage}
 \caption{Variation of $\psi$ in linear scale with $t$ for $q=0.2$, $\Lambda=0.0005$.}\label{dyn1}
 \end{figure}
where $\sigma_{1}$ and $\sigma_{2}$ represents width of the Gaussian packet and $\mu$ denotes position of the peak of the curves. It is found that broad Gaussian waves can not excite the background sufficiently to observe the dynamical features \cite{nils1999}. On the other hand, using sharp localised packet, one can get maximum number of extrema in field oscillation. First we have discretized the domain of integration $(r_*-t)$ plane by using
\begin{eqnarray}
x_i&=&x_0+i\Delta x,  i=0,1,2,3...\\
t_j&=&j\Delta t,     j=0,1,2,3...
\end{eqnarray}
Here $x$ is same as $r_*$. $\Delta t$ and $\Delta x$ are grid size of $y$ axis and $x$ axis respectively. $x_0$ is a point on boundary of $x$ axis.  
To determine the perturbation $\psi$ in advanced time, we use Taylor theorem in Eqn.(\ref{master_wave}) and get a discretized version of it: 
\begin{equation}
\psi_F=\frac{\Delta t^2}{\Delta x^2}(\psi_C-2\psi_E+\psi_A)+(2-\Delta t^2 V_E)\psi_E-\psi_D\label{advance},
\end{equation}
where, in general, F, C, E, A, D points are defined as : $\psi_F=\psi(x_i,t_{j+1})$,  $\psi_C=\psi(x_{i+1},t_j)$, $\psi_E=\psi(x_i,t_{j})$, $\psi_A=\psi(x_{i-1},t_{j})$, $\psi_D=\psi(x_i,t_{j-1})$, $V_E=V(x_i,t_{j})$.
 \begin{figure}[H]
 \begin{minipage}{.5\textwidth}
 \includegraphics[height=5 cm,width= 7 cm]{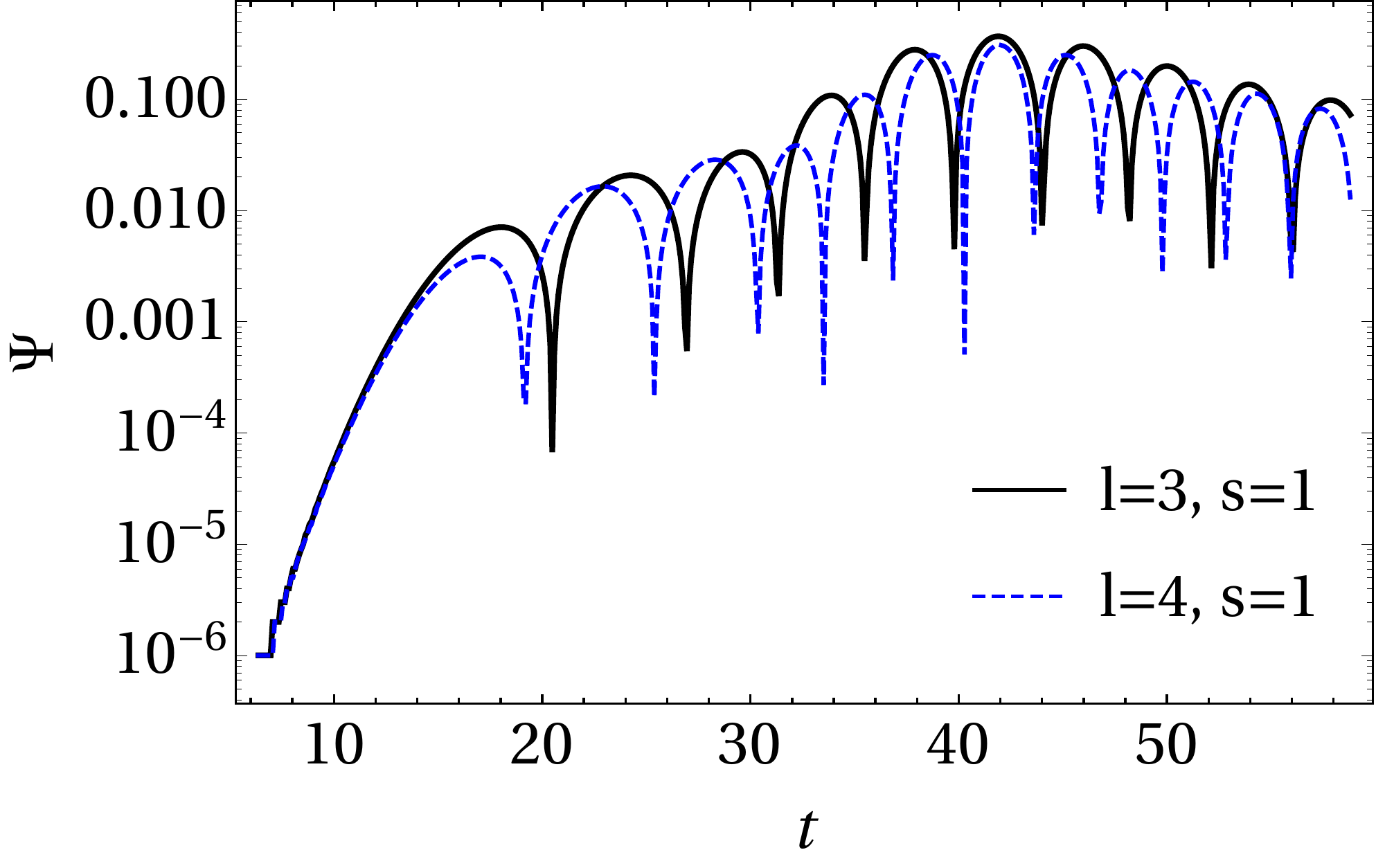}
 \end{minipage}%
  \begin{minipage}{.5\textwidth}
 \includegraphics[height=5 cm,width= 7 cm]{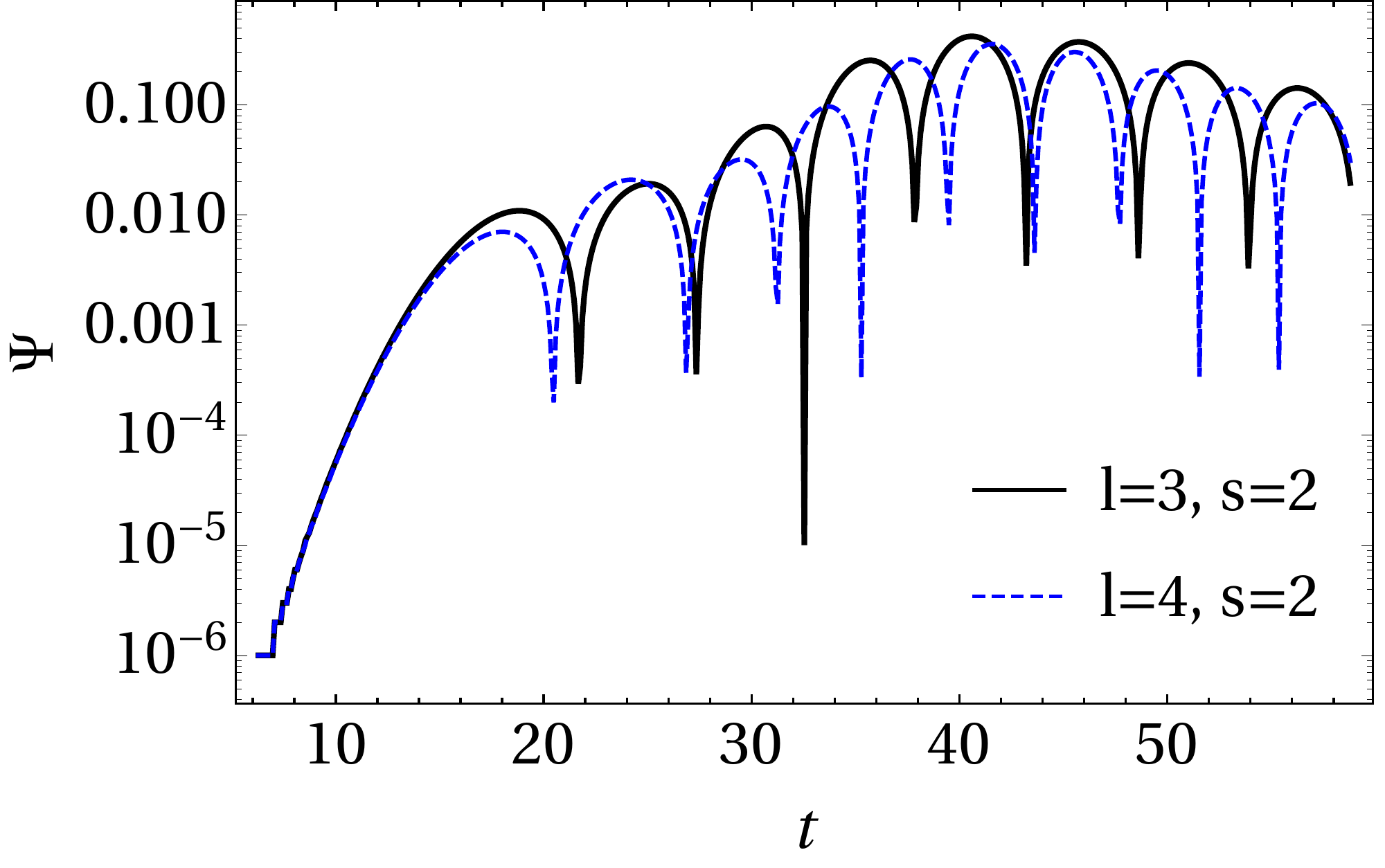}
 \end{minipage}
 \caption {Variation of $\psi$ in log scale with $t$ for $q=0.2$ and $\Lambda=0.0005$.}\label{dyn_log1}
 \end{figure}
With the initial conditions Eqn (\ref{int1}) and Eqn(\ref{int2}), we also specify all values of $\psi$ at $t=0$ and $t=\Delta t$ grid line of Fig.\ref{grid_final} . To determine the perturbation in one step advance (in time) at the point F, we need to know value of the perturbation in four neighbourhood points of F, which are represented by A, C, D and E. By applying this procedure repeatedly, one can determine the dynamics of perturbation over a complete domain. During this numerical integration scheme, one dimensional version of Courant-Friedrichs-Lewy (CFL) condition is satisfied which is a necessary condition for convergence of an explicit finite difference method of a hyperbolic PDE. Other parameters on which convergence depends are discussed in \cite{moli}. 
%
\begin{figure}[H]
  \centering
 \begin{minipage}{.5\textwidth}
  \centering
 \includegraphics[height=5 cm,width= 7 cm]{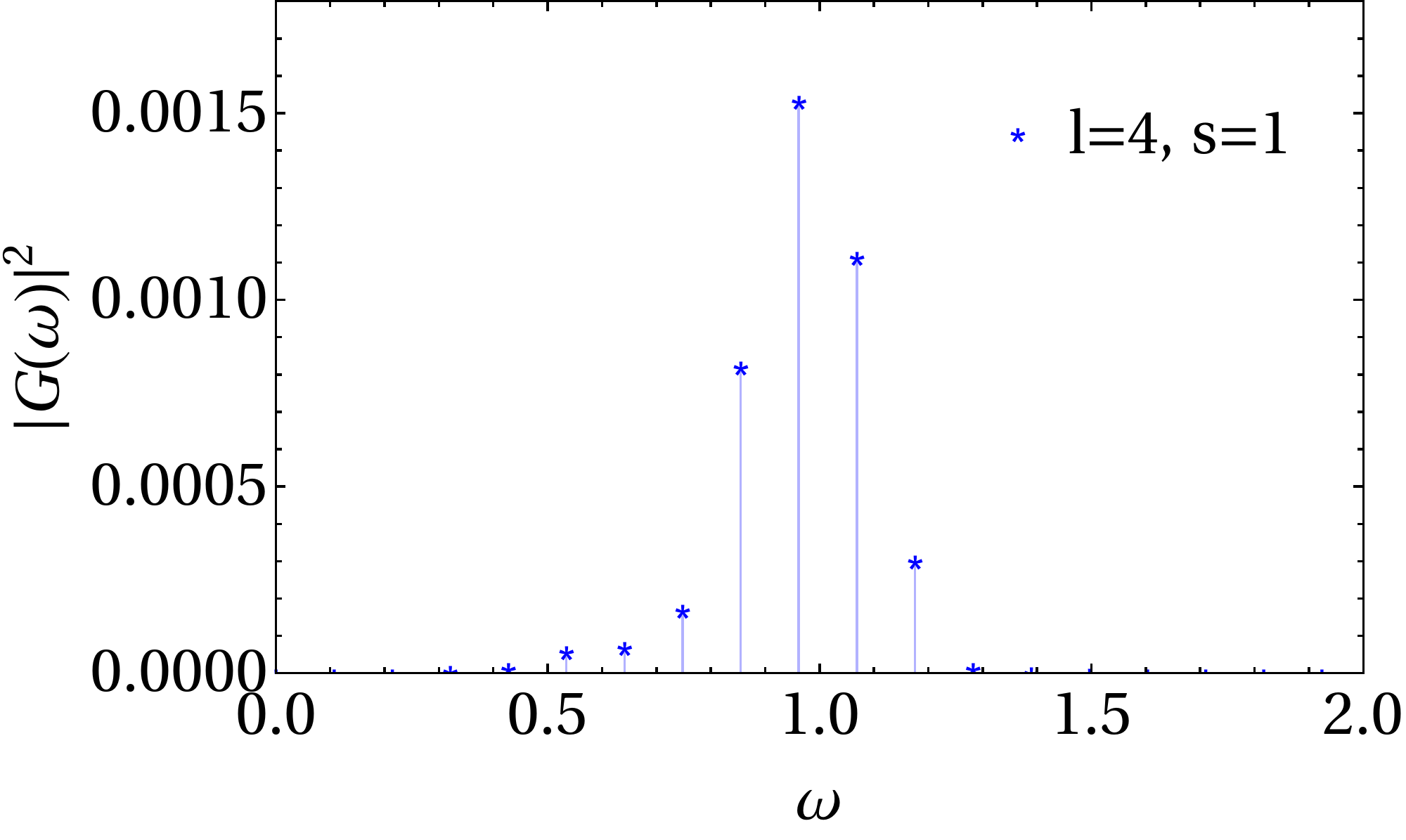}
 \end{minipage}%
  \begin{minipage}{.5\textwidth}
 \centering
 \includegraphics[height=5 cm,width= 7 cm]{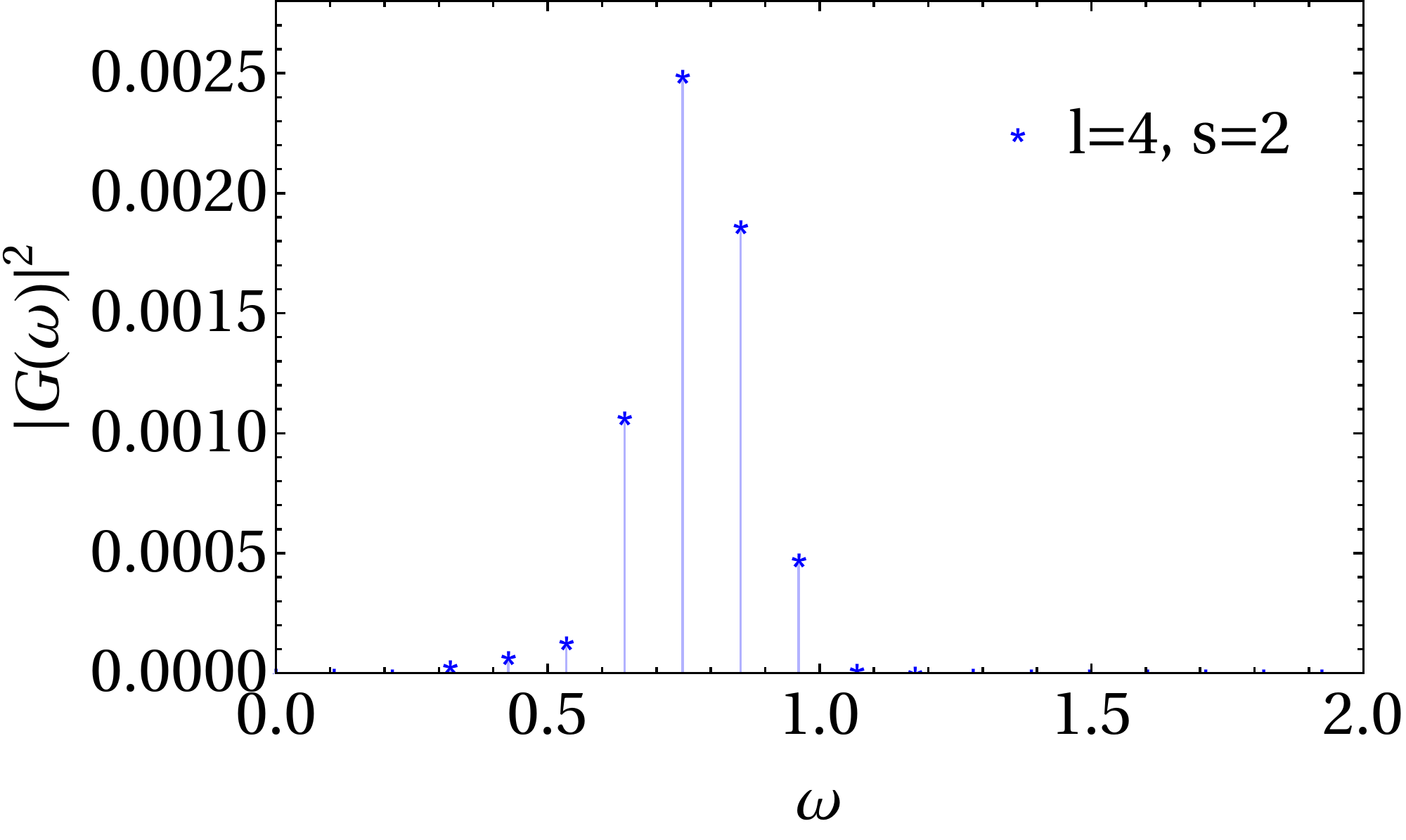}
 \end{minipage}
 \caption {Variation in power spectrum $|G(\omega)|^{2}$ with $\omega$ for $q=0.2$ and $\Lambda=0.0005$.}\label{fourier-trans}
 \end{figure}
In Fig.\ref{dyn1} we show the evolution of $\psi$ for two integral spin ($s=1$ and $s=2$ representing electromagnetic and gravitational perturbations respectively) perturbations with different $\ell$ values. It is quiet evident from plots that one of our motivations of studying stability feature of the scattered wave is fulfilled as it shows characteristic decay modes in late time. On the same time Fig.\ref{dyn_log1} exhibits evolution of perturbation in log scale. 
  It is clear from the applied numerical scheme that to find late time dynamics of $\psi$, we need initial conditions (\ref{int1}) , (\ref{int2}) on more number of grid points. As our integration domain is limited between finite values of $-r_*$ and $r_*$ because of finite value of cosmological horizon $r_c$, we were unable to generate numerical values of  $\psi$ at very late time and therefore any power law tail is absent in the dynamics. In real world, it is well known that $\Lambda$ is very small, of the order of $10 ^{-52}$. Use of this small value of $\Lambda$  in numerical computation has its own challenges. So we choose  a small, finite value for $\Lambda = 0.0005$ here, which, of course is not as small as the cosmological constant itself. To capture feature of small value of cosmological constant, one can decrease $\Lambda$ and recalculate it further.  Using Eqn.(\ref{metric}),  it is found that as  $\Lambda$ $\to$ $0$, $r_c \to \infty$. Therefore numerically domain of integration also becomes large enough and finally one can get very late time dynamics. 

 Fig.\ref{fourier-trans} is a power spectrum of Fourier transformation for the same perturbing wave of Fig.\ref{dyn1}.  Here we find independently oscillation frequency of perturbing wave from the frequency $\omega_0$ corresponding to maximum of $|G(\omega)|^{2}$. In $s=1$ and $s=2$  condition, maximum power contaning frequencies are $0.9617\pm0.1068$ and $0.7479\pm0.1068$ respectively. Scattered waves with these frequencies are dominant in Fig.\ref{dyn1}. Although there is no way to find specific overtone no$(n)$ of oscillation from the variation of $\psi$ but in late time it is expected that system will oscillate in fundamental mode \cite{abrahm}. By that time all higher modes will be damped out because of their large damping factors in frequencies (Table \ref{table1} -\ref{table2}). With $s=1$ and $s=2$, oscillation frequencies $(\omega_0)$ from the WKB method are $1.0398$ and $0.8151$ for $\ell=4$ and $n=0$ respectively. These oscillation frequencies are in good agreement with frequencies obtained from Fourier transformation technique.   

\section{Greybody factors and Absorption coefficients}
\subsection{Nature of the greybody factor}
In this subsection we will discuss Reflection coefficients $R(\omega)$ and Transmission coefficients $T(\omega)$ for different parameter spaces as well as in different type of perturbations. In \cite{fern2} coefficients for scalar type perturbation is discussed for Bds black holes in detail. Here, in order to fill the gap in the present literature, we will concentrate on the electromagnetic and gravitational perturbation part for this black hole. The use of WKB method to compute reflection and transmission coefficients(greybody factors) are not new. It has already been employed in various scenarios \cite{grain1}-\cite{nazir}, which includes the calculation of greybody factors of black holes in braneworld models, in the context of calculations of these coefficients for wormholes. By another analytical approach which was originally proposed by Unruh \cite{unruh}, greybody factors can also be calculated \cite{ciprian}. Here we have already seen that for both $s=1$ and $s=2$, finite potential barrier (Fig.\ref{plotep} , Fig.\ref{figure7} ) exists between cosmological horizon$(r_c)$ and event horizon$(r_h)$. Now,  any wave travelling past the cosmological horizon will face these finite positive potential barriers as obstacles. Therefore some part of the wave will be reflected back towards $r_c$ and some parts will be transmitted towards $r_h$. Following \cite{fern2}, we can represent them as
\begin{eqnarray}   
\psi(r_*)&=& T(\omega)e^{-i \omega r_*};   r_{*} \to  -\infty\label{bound_c1}  \\
\psi(r_*)&=& e^{-i \omega r_*}+R(\omega)e^{i \omega r_*};   r_{*} \to  +\infty  \label{bound_c2}
\end{eqnarray}
In general the reflection and transmission coefficients are functions of oscillation frequency $(\omega)$ of the wave. The reflection coefficient $R(\omega)$ in the WKB approximation is defined as,
\begin{equation}
R(\omega)={(1+e^{-2\pi i \alpha})}^{-\frac{1}{2}}\label{R_c},
\end{equation}
where $\alpha$ is given by
\begin{equation}
\alpha=\frac{i(\omega^2-V(r_0))}{\sqrt{-2V^{''}(r_0)}}-\Lambda_2-\Lambda_3\label{alpha},
\end{equation}
and expression for $\Lambda_2$, $\Lambda_3$ can be found out from Eqns.(\ref{part1}) and (\ref{part2}) respectively. Conservation of probability requires:
\begin{equation}
|R(\omega)|^2+|T(\omega)|^2=1\label{conserve}
\end{equation}
Finally, the greybody factor is defined as 
\begin{equation}
\gamma_\ell = |T(\omega)|^2\label{grey}
\end{equation}
Depending on the frequency and height of the potential barrier, there may be different cases which can arise: when $\omega^2 \gg V(r_0)$,  i.e. when a wave with frequency larger than the height of the barrier comes, it will not be reflected by the barrier classically. In this case, one should expect the reflection coefficient to be close to zero, because the frequency of the wave is large enough to cross the barrier. Therefore we expect that under this conditions,  $|T|^2$ will be close to $1$. When $\omega^2 \ll V(r_0)$, i.e. square of the frequency is very small compared to barrier height, wave will be reflected back from the barrier and some part may be transmitted through the barrier by tunnelling effect depending on the values of $\omega$ and $V(r_0)$. We should get exactly opposite behaviour of $R(\omega)$ and  $T(\omega)$ compared to previous case. In this case, the WKB method does not have very high accuracy.  When $\omega^2 \approx V(r_0)$, we have to take help of numerical techniques to understand the nature of  $R(\omega)$ and  $T(\omega)$. Here we can apply WKB approximation method because of the small distance between the turning points. 

\begin{figure}[H]
	\begin{minipage}{.5\textwidth}
		\includegraphics[height=5 cm,width= 7 cm]{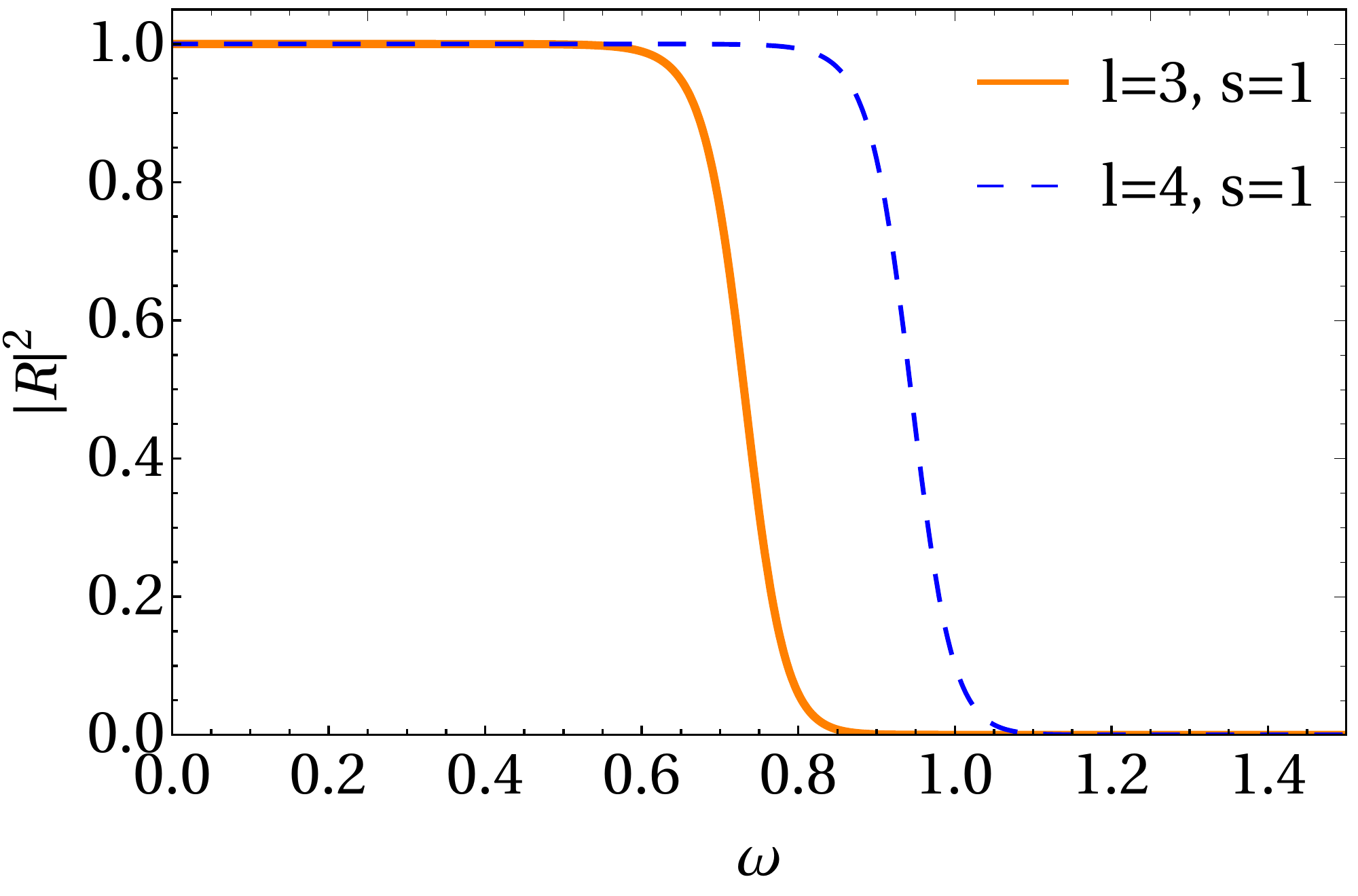}
	\end{minipage}%
	\begin{minipage}{.5\textwidth}
		\includegraphics[height=5 cm,width= 7 cm]{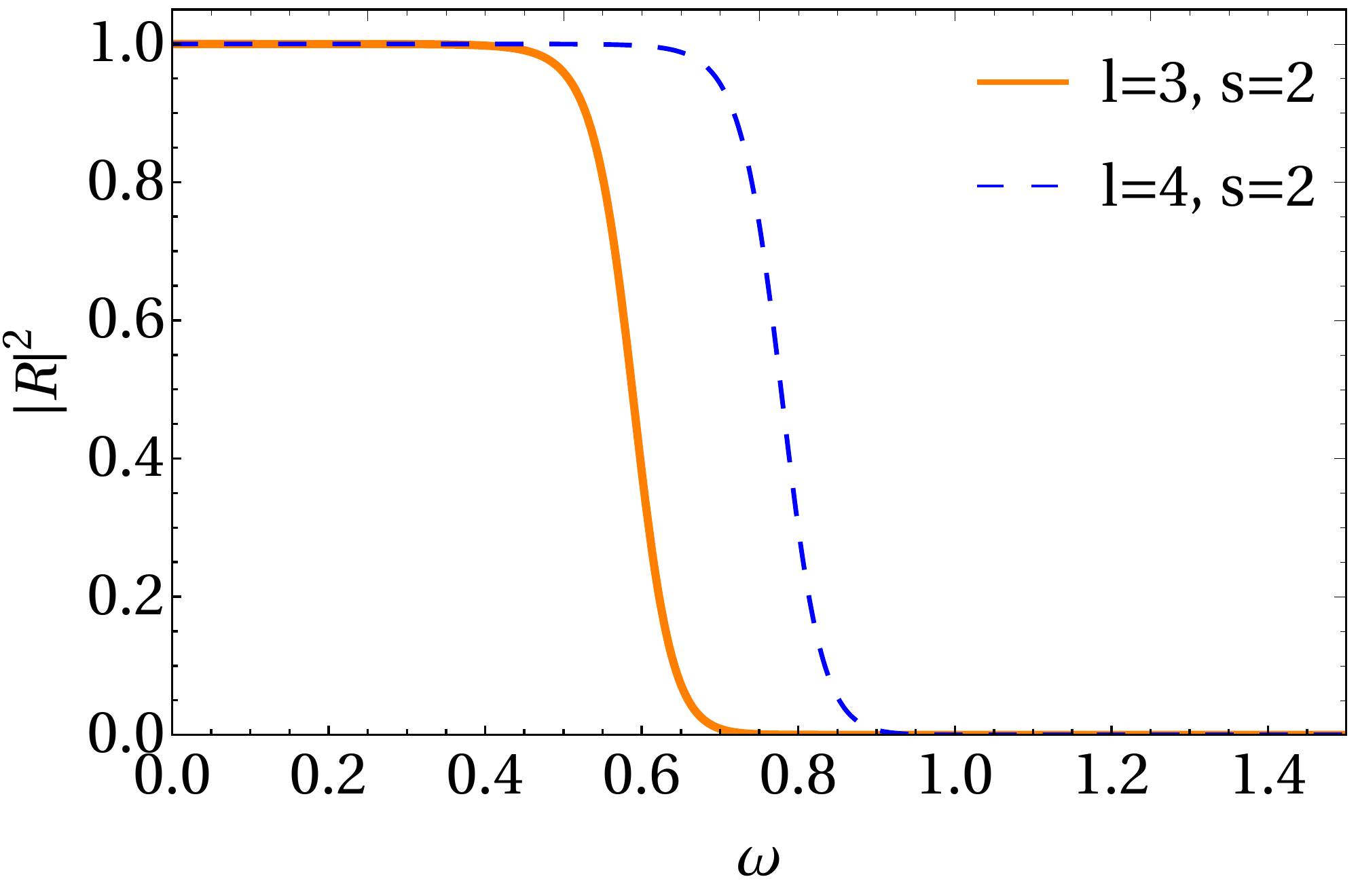}
	\end{minipage}
	\caption {$|R|^2$ vs $\omega$ for $q=0.4$, $\Lambda=0.02$.}\label{R_l_1}
\end{figure}

\begin{figure}[H]
	\begin{minipage}{.5\textwidth}
		\includegraphics[height=5 cm,width= 7 cm]{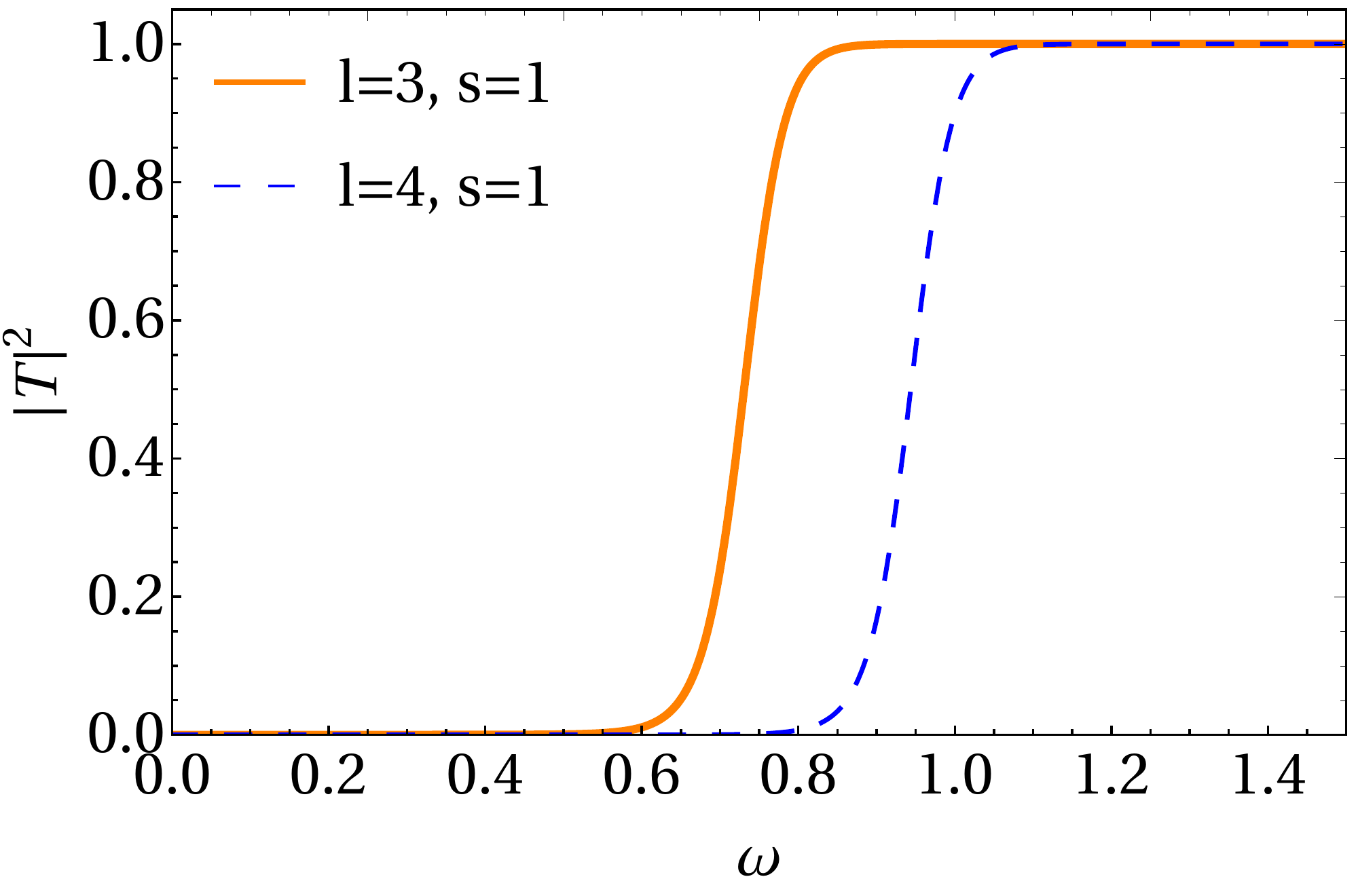}
	\end{minipage}%
	\begin{minipage}{.5\textwidth}
		\includegraphics[height=5 cm,width= 7 cm]{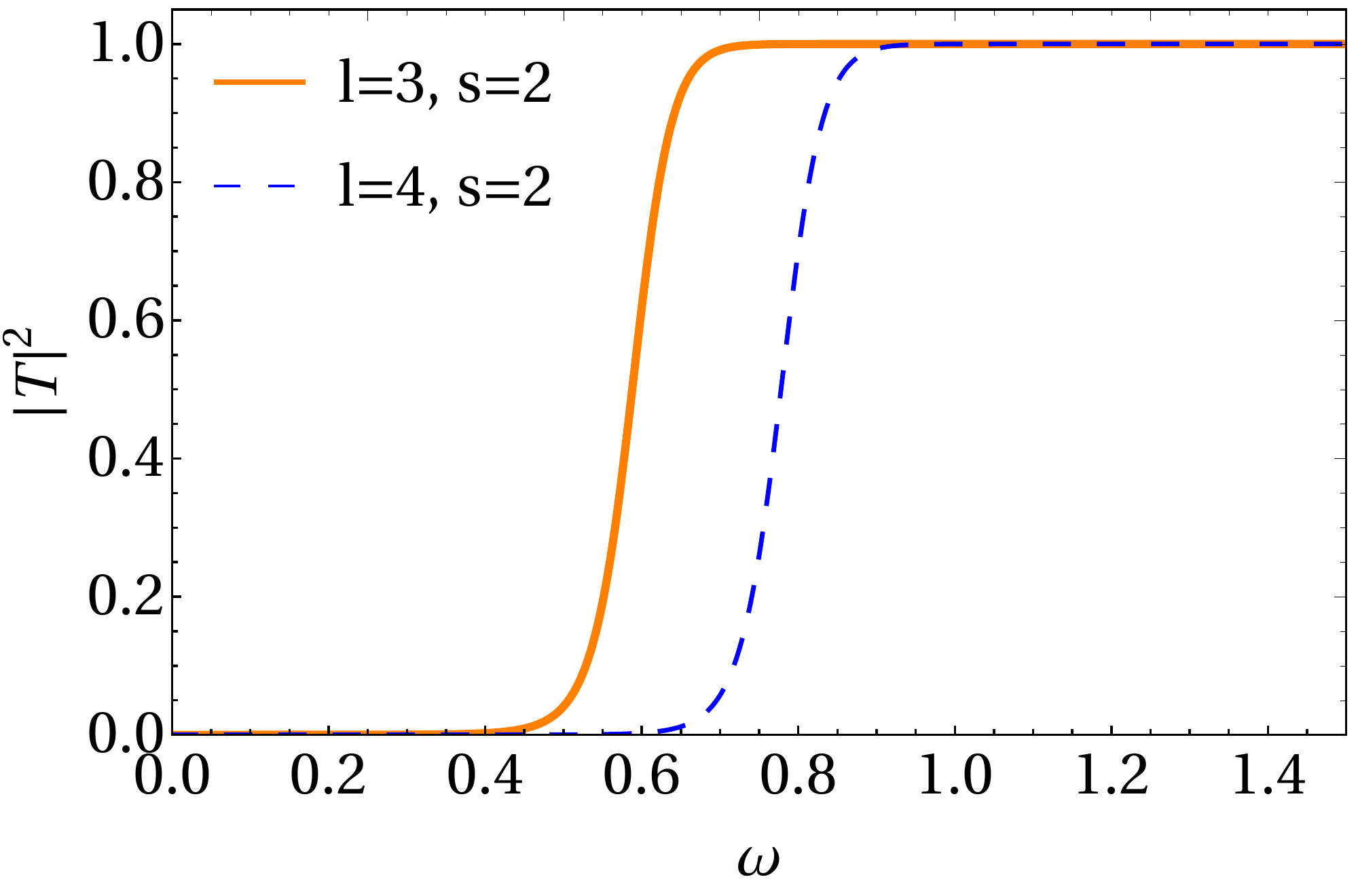}
	\end{minipage}
	\caption {$|T|^2$ vs $\omega$ for $q=0.4$, $\Lambda=0.02$.}\label{T_l_1}
\end{figure}

\begin{figure}[H]
	\centering
	\begin{minipage}{.5\textwidth}
		\includegraphics[height=5 cm,width= 7 cm]{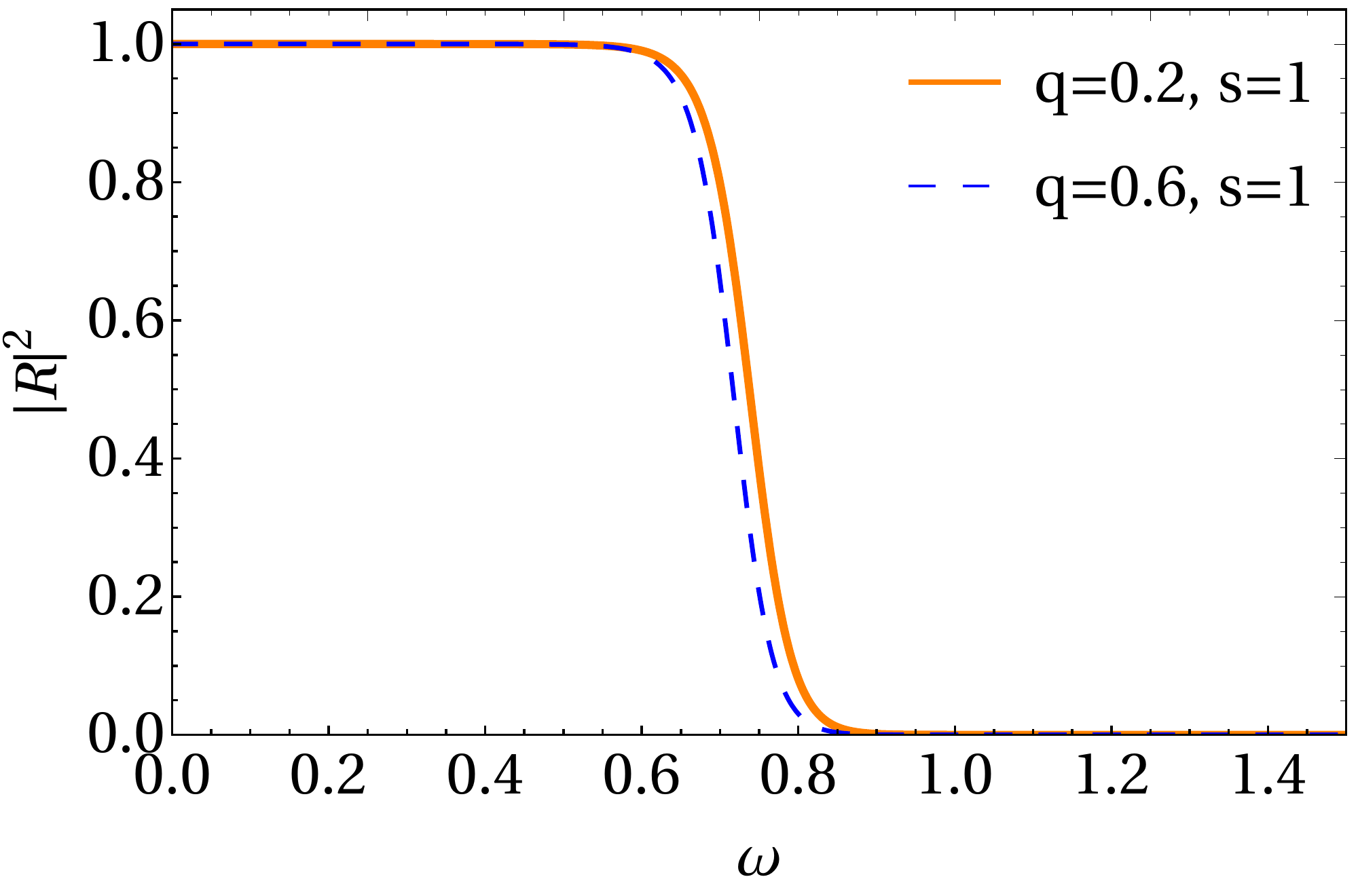}
	\end{minipage}%
	\begin{minipage}{.5\textwidth}
		\includegraphics[height=5 cm,width= 7 cm]{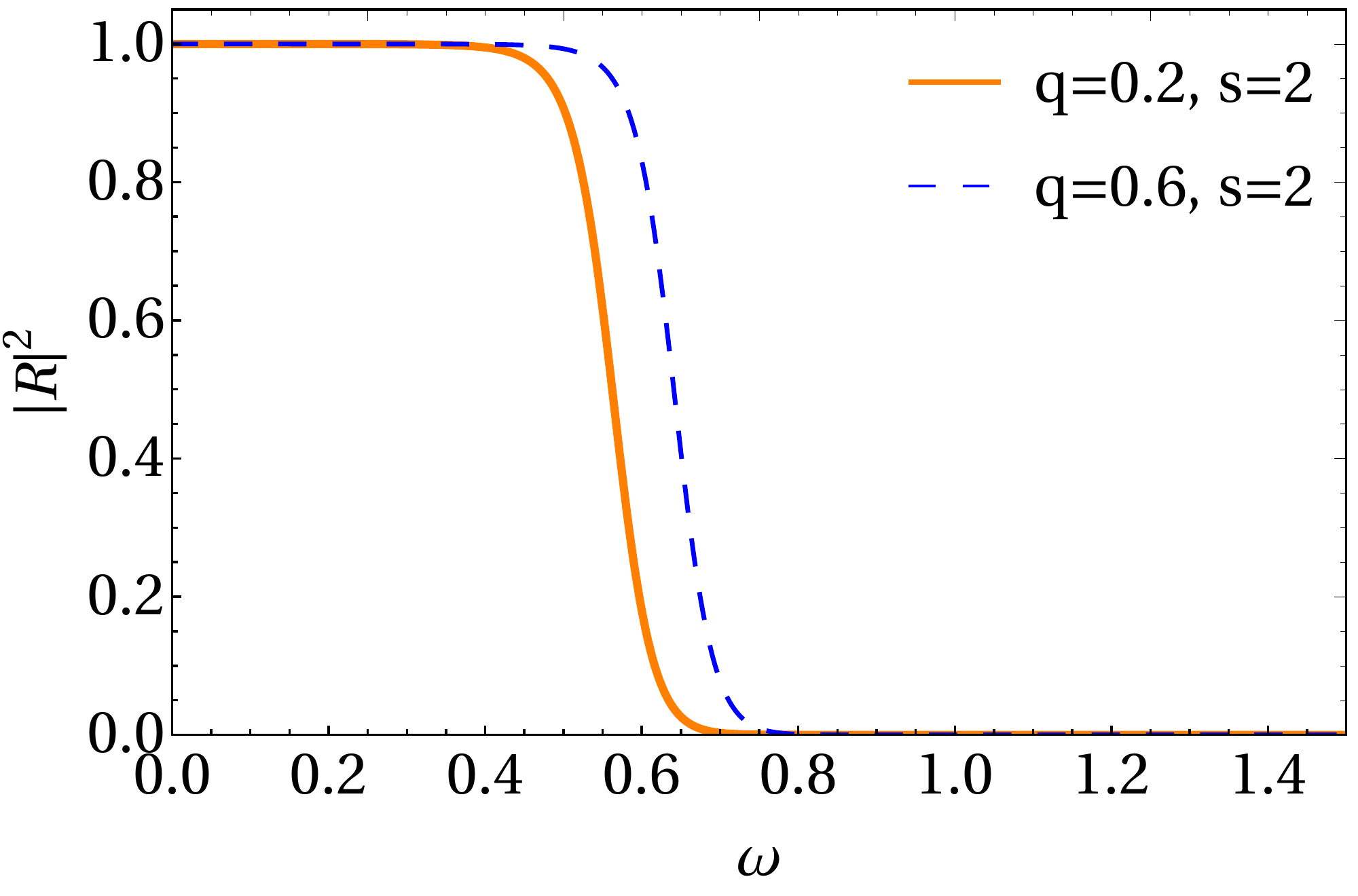}
	\end{minipage}
	\caption {$|R|^2$ vs $\omega$ with $\ell =3$ and $\Lambda=0.02$ for different magnetic charge ($q$) values.}\label{r_q}
\end{figure}

\begin{figure}[H]
	\begin{minipage}{.5\textwidth}
		\includegraphics[height=5 cm,width= 7 cm]{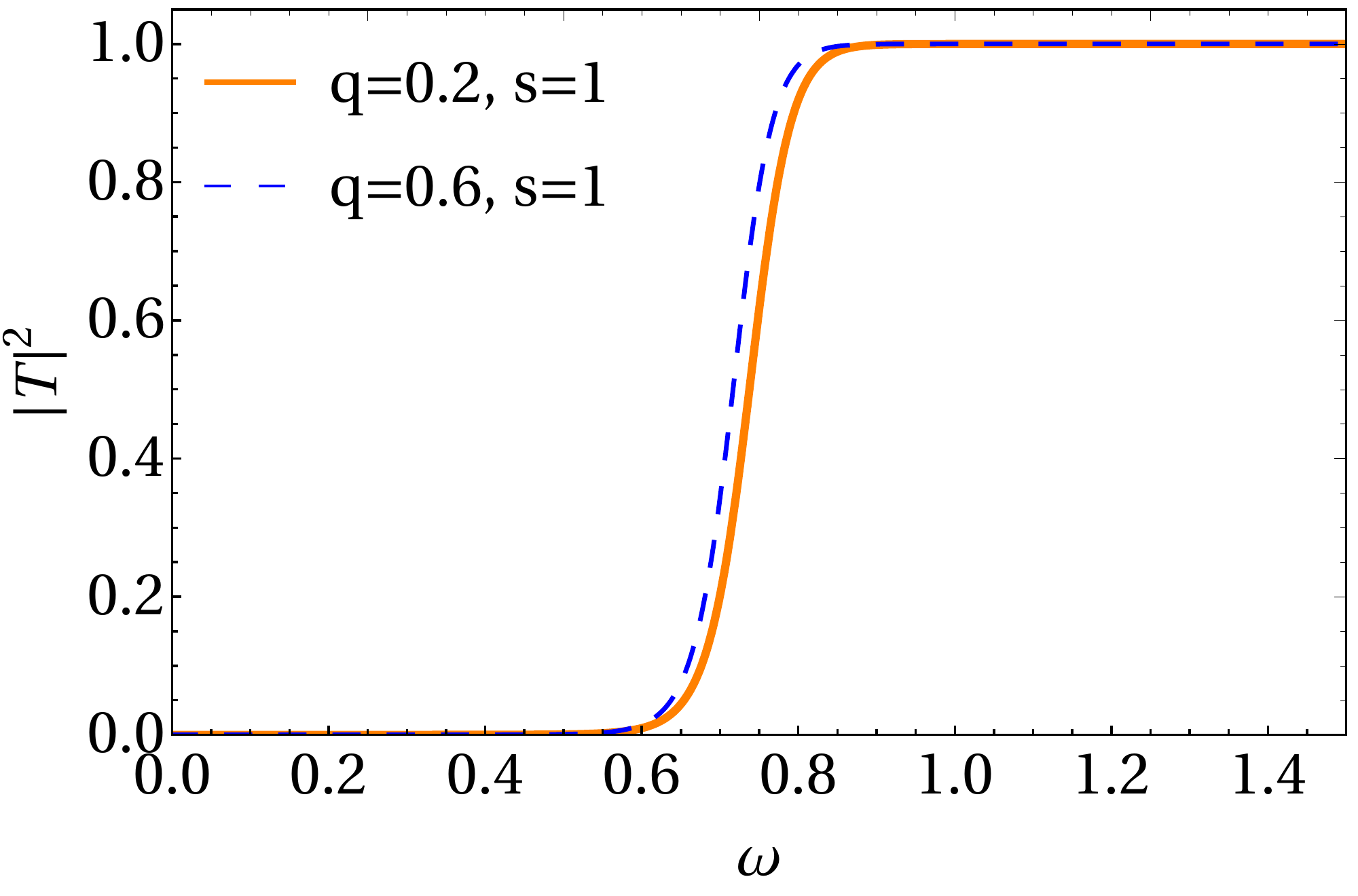}
	\end{minipage}%
	\begin{minipage}{.5\textwidth}
		\includegraphics[height=5 cm,width= 7 cm]{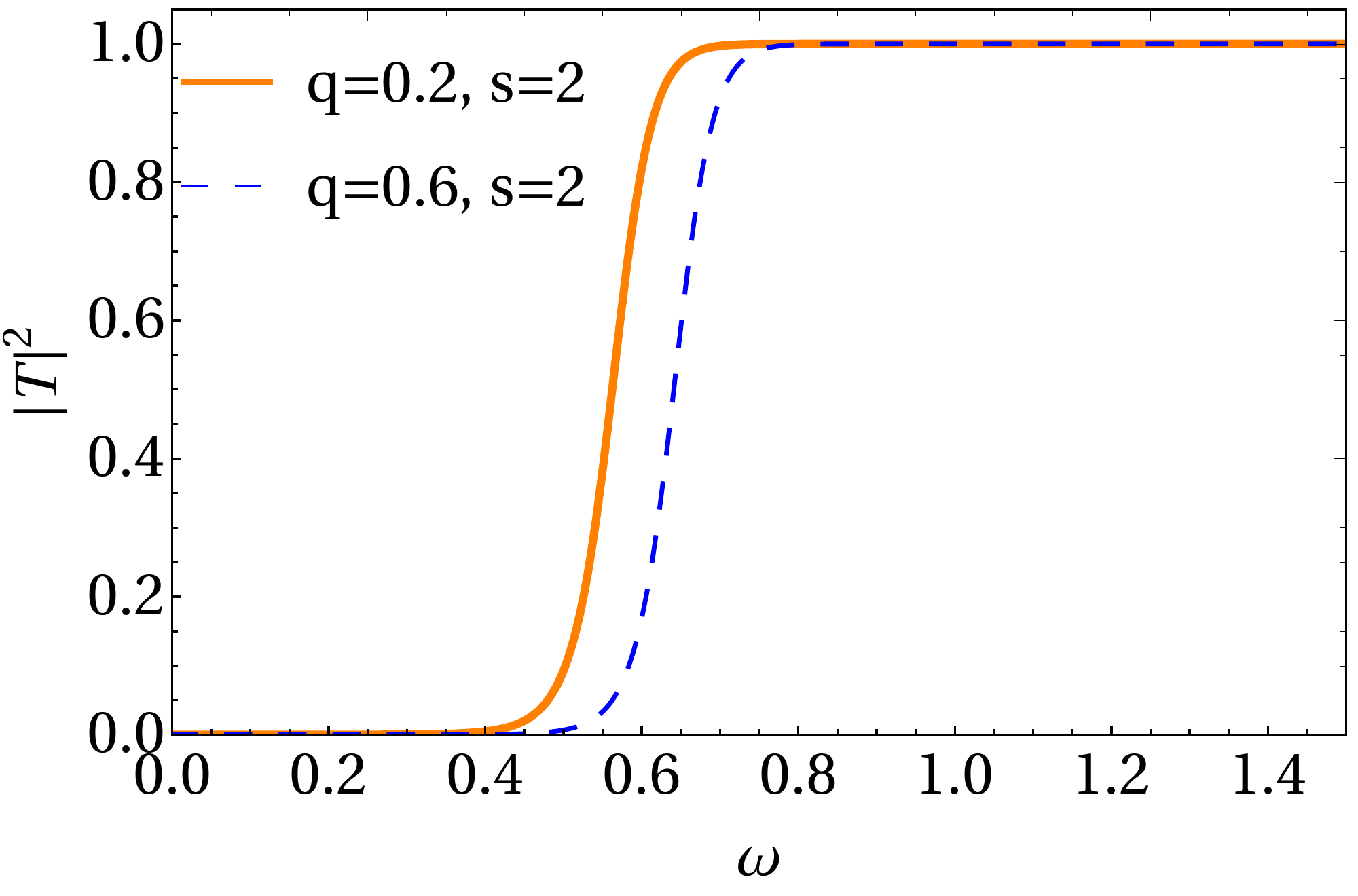}
	\end{minipage}
	\caption {$|T|^2$ vs $\omega$ with $\ell =3$ and $\Lambda=0.02$ for different magnetic charge ($q$) values.}\label{T_q}
\end{figure}

\begin{figure}[H]
	\centering
	\begin{minipage}{.5\textwidth}
		\includegraphics[height=5 cm,width= 7 cm]{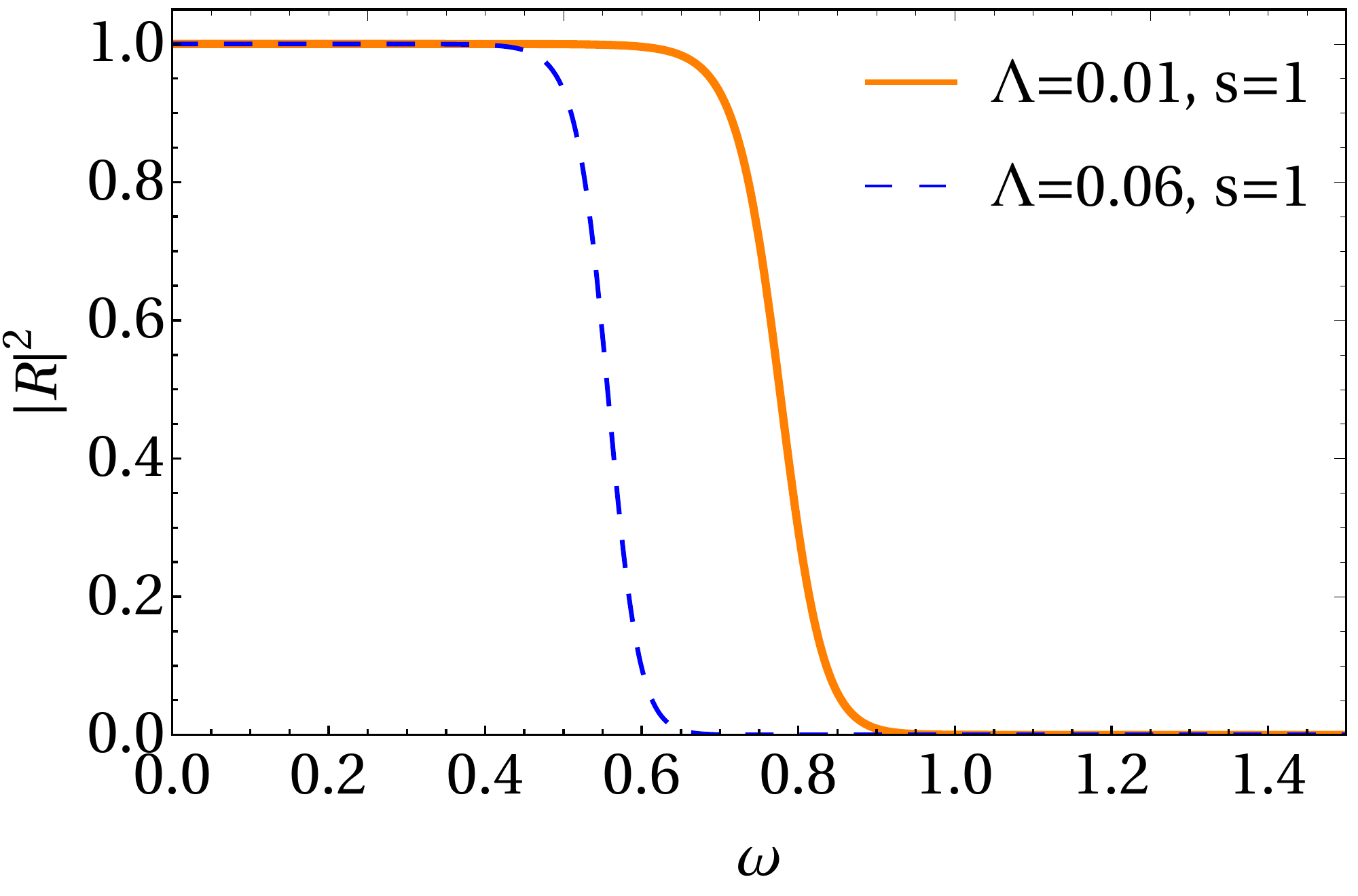}
	\end{minipage}%
	\begin{minipage}{.5\textwidth}
		\includegraphics[height=5 cm,width= 7 cm]
{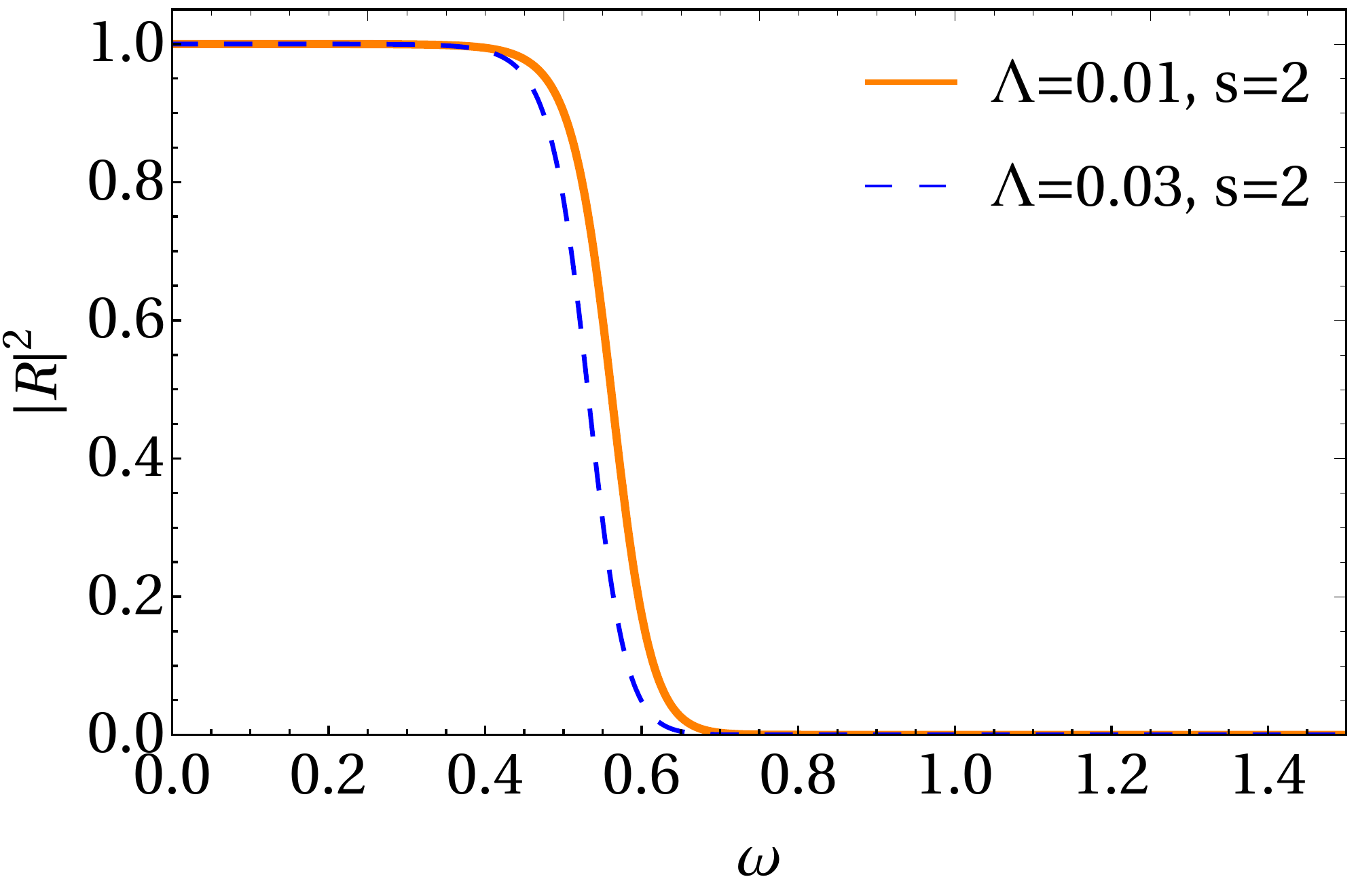}
	\end{minipage}
	\caption {$|R|^2$ vs $\omega$ for $\ell =3$ and $q=0.2$ for different $\Lambda$ values.}\label{r_lamb}
\end{figure}

\begin{figure}[H]
	\centering
	\begin{minipage}{.5\textwidth}
		\includegraphics[height=5 cm,width= 7 cm]{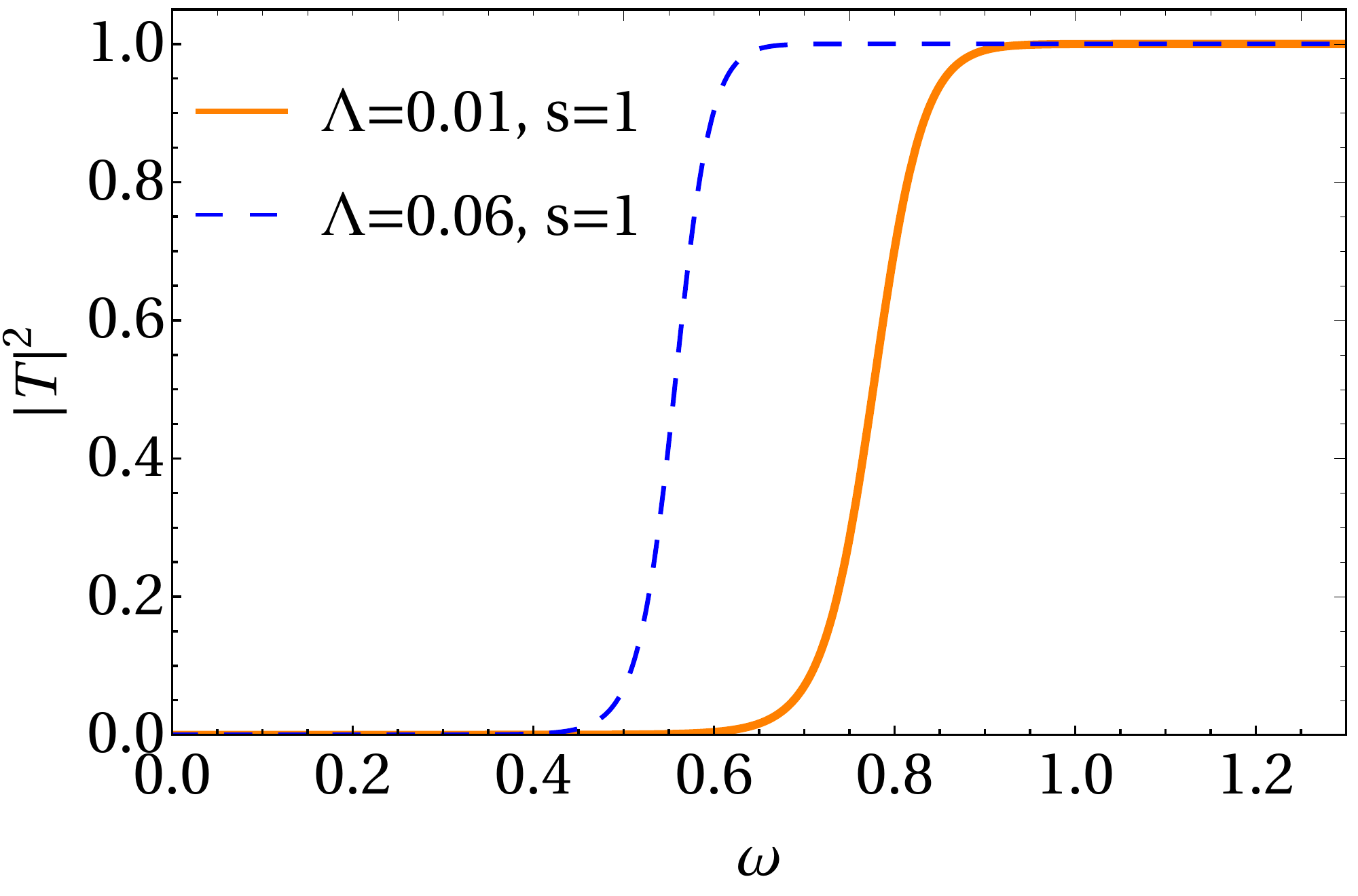}
	\end{minipage}%
	\begin{minipage}{.5\textwidth}
		\includegraphics[height=5 cm,width= 7 cm]{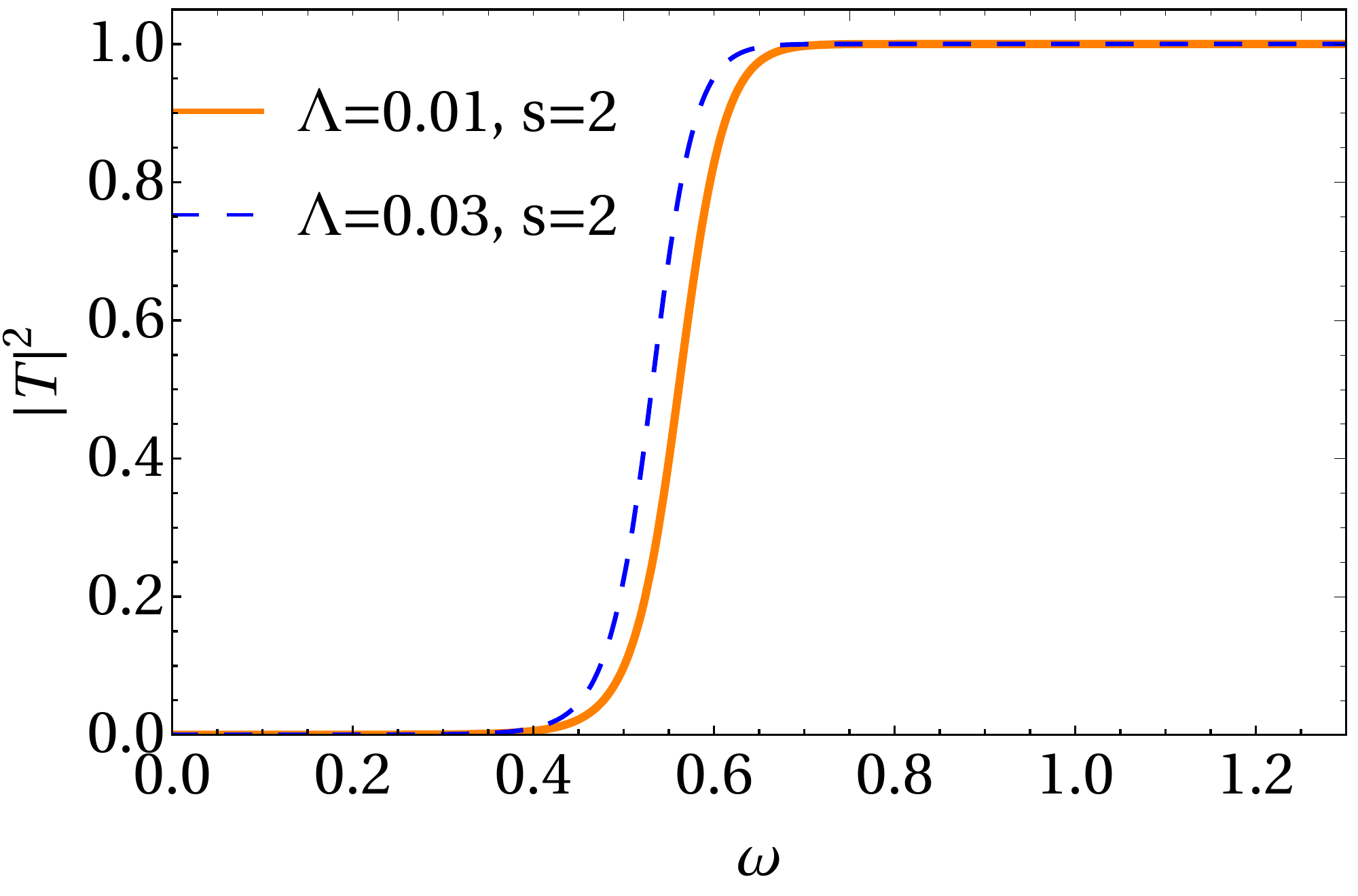}
	\end{minipage}
	\caption {$|T|^2$ vs $\omega$ for $\ell =3$ and $q=0.2$ for different $\Lambda$ values.} \label{t_lamb}
\end{figure}

Fig.\ref{R_l_1} shows variation of $|R(\omega)|^2$  with $\omega$ for different class of perturbations with different $\ell$ values. In electromagnetic perturbations, i.e.$(s=1)$, it is almost one for low frequency and for high frequency it is close to zero. For a fixed frequency, $|R|^2$ is larger for multipole number $\ell =4$ than $\ell =3 $. It can be explained easily from the dependency of effective potential $V(r)$ on $\ell$. Gravitational perturbation $(s=2)$ has also same nature and features like that of the electromagnetic type $(s=1)$. On the contrary Fig.\ref{T_l_1} shows variation of $|T(\omega)|^2$  with $\omega$ for both $(s=1)$ and $(s=2)$ type. Following Eqn.(\ref{conserve}), it shows exactly opposite nature to Fig.\ref{R_l_1} for both the limits.
Next we study the behaviour of $|R(\omega)|^2$  with $\omega$ by varying black hole magnetic charge $q$ and keeping other parameters fixed in Fig.\ref{r_q} . Larger $q$ values decreases reflection coefficient compared to smaller $q$ values for $s=1$. Although for $s=2$, response of $|R(\omega)|^2$ under different $q$ is much larger than $s=1$. In gravitational perturbation, for higher charge parameter, reflection coefficient is also larger which is exactly opposite to electromagnetic perturbation. This nature is prominent from the potential behaviour of different class of perturbation under variation of charge parameter. Fig.\ref{T_q} shows $|T(\omega)|^2$  with $\omega$ with the same parameter values.
Next we plot $|R(\omega)|^2$  vs $\omega$ by varying the cosmological constant $\Lambda$ in Fig.\ref{r_lamb} . For $s=1$, response of $|R(\omega)|^2$ under different $q$ is much larger than $s=2$. With increasing $\Lambda$ for both $s=1$ and $s=2$, $|R(\omega)|^2$ value decreases. Therefore in space time with larger cosmological constant, black holes can less scatter the incoming waves. Fig.\ref{t_lamb} shows the variation of $|T(\omega)|^2$  with $\omega$ with the same set of parameter values following Eqn. (\ref{conserve}).
\subsection{Absorption Cross-section}
\begin{figure}[H]
		\centering
		\includegraphics[height=5 cm,width= 7 cm]{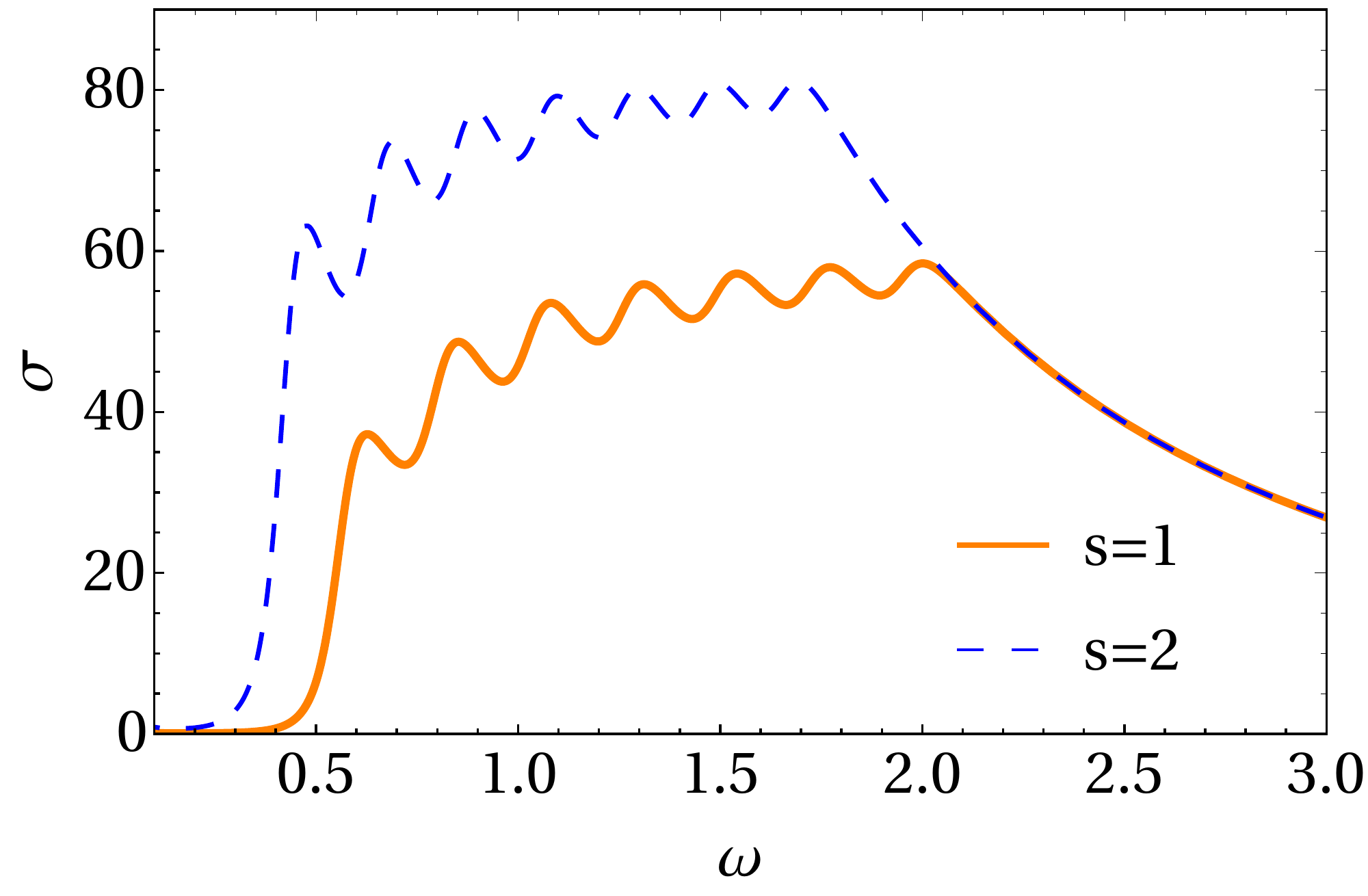}
	\caption {Total absorption cross section($\sigma$) vs $\omega$ for 
		$q=0.4$ and $\Lambda=0.002$}\label{cross-sec}
\end{figure}

In this subsection we will discuss partial and total absorption cross section in the context of electromagnetic and gravitational perturbation for different parameter spaces in the BdS background. Partial $(\sigma_\ell)$ and total absorption cross sections $(\sigma)$ are defined respectively as:  
\begin{eqnarray}
\sigma_{\ell}&=& \frac{\pi(2\ell+1)}{\omega^2}|T_{\ell}(\omega)|^2\label{par_cross},\\
\sigma&=&\displaystyle\sum_{\ell} \frac{\pi(2\ell+1)}{\omega^2}|T_{\ell}(\omega)|^2\label{crs-sec-eq}.
\end{eqnarray}
In Fig.\ref{cross-sec}, variation of $\sigma$ are plotted with different $q$ and $\Lambda$ values where individual peak represents $\sigma_{\ell}$. For both $s=1$ and $s=2$, Total absorption cross sections have similar feature.
Variation of $\sigma$ in Fig.\ref{cross-sec} can be classified into three distinct regions, the first region consists of a growing phase which is the signature of increasing $|T(\omega)|$ with $\omega$. The second region shows oscillations in $\sigma$ which comes considering different $\ell$ modes. In this particular example, we have added upto $\ell=8$ modes to determine $\sigma$. The last part is a power law fall-off. The reason behind this fall-off is following: after certain critical frequency $(\omega_0)$, the transmission coefficient attains maximum value to $1$. Afterwards, with further increase in $\omega$, $\sigma$ becomes proportional to $\frac{1}{\omega^2}$ irrespective of the type of the perturbation or the values of the parameters of the black hole space-time. The electromagnetic ($s=1$) and the gravitational ($s=2$) part shows two different branches which merges on top of each other in the fall-off region. For a fixed frequency, absorption cross section is always larger for gravitational perturbation w.r.t electromagnetic one.

\section{Summary and Conclusion}

In this paper, we have focused on two most important types of black hole perturbations: electromagnetic and gravitational, for a regular BdS black hole.  We have used sixth order WKB approximation method to compute the QN frequencies of the BdS black hole under these perturbations and found out the response of the black hole to these perturbations by varying different parameters of the space-time. It is easy to see that from one type to another type of perturbation, only the potential profile changes in the Schr\"{o}dinger-like wave equations, while keeping forms of all the relevant equations intact. We studied how the frequencies vary  as a function of multipole number ($\ell$) as well as with the parameters like the cosmological constant ($\Lambda$), magnetic charge  ($q$) and overtone number $(n)$. 
As the multipole number ($\ell$) increases, both Re ($\omega$) and -Im ($\omega$) increase for a fixed overtone number ($n$).  The real oscillation frequency and imaginary part of the frequency representing damping are decreasing and increasing respectively with increasing overtone number $n$ for fixed $\ell$ values for the axial EM perturbations. It was observed that the real part of the QN frequency increases monotonically with the multipole number whereas the imaginary part at the beginning starts to increase but then it saturates after reaching for a certain $\ell$ value. While for the imaginary part of the gravitational QN frequency, the frequency initially increases and then falls down before finally saturation occurs. This behaviour of imaginary part being constant with the variation of multipole number is a common feature of  both the electromagnetic as well as gravitational perturbations, although the rapidity with which the imaginary part of the frequency increases is more in the case of electromagnetic one as compared with gravitational perturbations. We have also conducted a study to find out the Q-factor of the BdS black hole system and found that through Q-factor one can differentiate between electromagnetic and  gravitational perturbation. The Q-factor increases rapidly with charge $q$ and falls down after a critical value of the charge, while it decreases slightly with the increase of $\Lambda$ for electromagnetic perturbations. It increases non-linearly with $q$ and decreases very slowly with $\Lambda$ for gravitational perturbation. In both the cases of electromagnetic and gravitational perturbations, the negative imaginary part shows similar behaviour with increasing $q$, namely, they decrease with an increase in the magnetic charge. However, for the real part, although it decreases with charge for EM perturbations, opposite behaviour is found for the gravitational perturbations. In all these cases, we did a comparative study with respect to the RN-dS background also and showed the effect of non-linear electrodynamics on the nature of QN frequencies when studied with respect to different parameters. We have further studied the dynamics of the perturbations using a standardised numerical integration method. Finally, we investigate the reflection and transmission coefficients from the BdS black hole due to electromagnetic and gravitational perturbations. In both the cases, behaviour of the greybody factor were studied by varying the black hole parameters. The total absorption cross section for different multipole values (upto $\ell=8$) was studied. 

For future directions, it would be interesting to study whether isospectrality of the QN spectrum holds in BdS spacetime or not. It is already known that both the axial and polar perturbations (electromagnetic as well as gravitational) gives rise to same QN frequencies. The well known example being the Reissner-Nordstr\"{o}m black hole arising out of Einstein's general theory of relativity coupled to Maxwell's electrodynamics. However, to our knowledge, no such works in case of regular black holes in de Sitter space time exist. It would therefore be interesting to study the isospectrality of different types of perturbations in regular black holes in de Sitter space. Another important area of study would be to look at the circular null geodesics of the near extremal BdS black holes and find out whether there can be any relation between the Lyapunov exponents and the QNMs of the BdS background.   Finally we would like to mention that, although many works have been done on the electromagnetic and gravitational perturbations of black holes in the regime of Einstein's general theory of relativity, not many examples are present in the context of regular black holes. Particularly, the stability and QN properties of regular black holes in de Sitter universe remains a very less studied area in the literature so far. We believe that this work will fill the gap.

\begin{appendices}
  \section{Appendix: The gravitational perturbation}

In this appendix we will briefly discuss about the gravitational perturbation in general. The spherically symmetric, static background metric is represented by $g^{0}_{\mu\nu}$ and the small perturbation to the background metric is denoted by $h_{\mu\nu}$. In order to perform the calculation to linearise the Einstein equation, we follow $|h_{\mu \nu}| \ll 1 $. Then $R_{\mu\nu}$  is evaluated from $g^{0}_{\mu\nu}$ and $R_{\mu\nu}+ \delta R{\mu\nu}$ from $g_{\mu\nu}=g^{0}_{\mu\nu}+h_{\mu\nu}$.
\begin{equation}
\delta R_{\mu\nu}=\delta\Gamma^{\alpha}_{\ \mu\alpha;\nu}-\delta\Gamma^{\alpha}_{\ \mu\nu;\alpha}\label{delR}
\end{equation}
where 
\begin{equation}
\delta\Gamma^k_{\ \mu\nu}=\frac{1}{2}g^{k\alpha}(h_{\alpha\nu;\mu}+h_{\alpha\mu;\nu}-h_{\mu\nu;\alpha})
\end{equation}
Now using Regge-Wheeler gauge for Axial type perturbation, the canonical form for the perturbation takes \cite{nol, rg}
\[h_{\mu\nu}=
\displaystyle\sum_{\ell=0}^{\infty} \displaystyle\sum_{m=-\ell}^{\ell}\begin{pmatrix}

0 & 0 & -h_0(t,r)\frac{1}{\sin\theta}\frac{\partial Y_{lm}}{\partial\phi} &  h_0(t,r)\sin\theta \frac{\partial Y_{lm}}{\partial\theta} \\ 

 & & & \\ 

* & 0 &  -h_1(t,r)\frac{1}{\sin\theta} \frac{\partial Y_{lm}}{\partial\phi}  & h_1(t,r) \sin\theta \frac{\partial Y_{lm}}{\partial\theta} \\

 & & & \\ 
 
* & * & 0 & 0\\
 
  & & & \\ 

* & * & * & 0

\end{pmatrix}
\]

where, $*$ marked components of the metric are determined by symmetry property of $h_{\mu\nu}$. Now we substitute the total metric $g_{\mu\nu}$ to the left side of Eqn.(\ref{Einfulll}) and calculate right hand side using Eqn.(\ref{ATotal}). Here we have taken into account the perturbation in the energy–momentum tensor and get linearized Einstein's equation. 
\begin{equation}\label{Einfulll}
	G_{\mu\nu}+\Lambda g_{\mu\nu}=2\left(\frac{\partial \mathcal{L}(F)}{\partial F}F_{\mu\lambda}F^{\lambda}_{\nu}-g_{\mu\nu}\mathcal{L}(F)\right)
\end{equation}
 We get coupled second order partial differential equations from $r\phi,\theta\phi$ and $t\phi$ components of Einstein's equation as Eqn.(\ref{pert1}), (\ref{pert2}) and (\ref{pert3}) respectively. These equations  are generalised equations for evolution of axial gravitational perturbation in Regge-Wheeler gauge. They are derived without any loss of generality.

\begin{eqnarray}
&&\left(\frac{\partial^2 h_{1}}{\partial t^2}-\frac{\partial^2 h_{0}}{\partial t\partial r}+\frac{2}{r}\frac{\partial h_0}{\partial t}\right)+\frac{f h_1}{r^2} \left(K+2 r^2 (2\mathcal{L}+\Lambda)+r \left(f'+\left(rf'\right)'\right)\right)= 0 \label{pert1}\\
&&\frac{\partial h_0}{\partial t}-f\frac{\partial \left(h_1 f\right)}{\partial r}=0\label{pert2} \\
&&\left(\frac{\partial ^2 h_0}{\partial r^2}-\frac{\partial ^2 h_1}{\partial t \partial r}-\frac{2}{r} \frac{\partial h_1}{\partial t} \right)-\frac{h_0}{f r^2}\left(K+2 r^2 (2\mathcal{L}+\Lambda)+2f+r \left(f'+\left(rf'\right)'\right)\right)=0\label{pert3}
\end{eqnarray}
Here $K=(\ell-1)(\ell +2).$ Next, as a standard method, one defines 
\begin{equation}
Q(t,r)=\frac{f(r)h_1(t,r)}{r},
\end{equation}
and after substituting $\frac{\partial h_0(t,r)}{\partial t} $ from Eqn. (\ref{pert2}) to  Eqn. (\ref{pert1}), we get the Schr\"{o}dinger-like equation and the generalized effective potential for the gravitational perturbation, which is denoted by $V$.
\begin{equation}
\frac{\partial ^2 Q(t,r_*)}{\partial t^2}-\frac{\partial ^2 Q(t,r_*)}{\partial r^2 _*}+Q(t,r_*)V(r)=0 \label{grav_wave},
\end{equation}
where,
\begin{equation}\label{gravpot}
V(r)=f \left[\frac {\ell(\ell+1)+r \left(rf'\right)'+2(f-1)+2 r^2 (2\mathcal{L}+\Lambda)} {r^2}\right]
\end{equation}
 Here $'$ denotes derivative with respect to the radial coordinate r. In \cite{bobirgr}, Eqn. (\ref{pert1}), (\ref{pert2}) and (\ref{pert3}) are derived for asymptotically flat space time. They are same as our set of equations and finally the effective potential due to gravitational perturbation turns out to be also same as was found in Eqn.(\ref{gravpot}) in the limit $\Lambda=0$. However, there is a difference in numerical factor (coefficint of $\mathcal{L}$) in the  effective potential, which is an artefact of the two different coefficients of $\mathcal{L}$ in our (\ref{action}) compared to the one used in \cite{bobirgr}.   
 
\end{appendices}


{\bf{Acknowledgement}}

\vspace{.5cm}

\noindent
SD would like to thank Nilanjandev Bhaumik, Sambrito Ghatak, Ranjini Mandal and Andrea Maselli for discussions on the numerical solution of PDE, CFL condition and technique of computing first order perturbations in the domain of General Relativity.  

\bibliographystyle{unsrt}

\end{document}